\documentclass[aps,prb,twocolumn]{revtex4}
\usepackage[latin1]{inputenc}
\usepackage{amsmath}
\usepackage{amssymb}
\usepackage{graphicx}

\begin{document}

\title{Out-of-plane fluctuation conductivity of layered superconductors
in strong electric fields}

\author{I. Puica}

\author{W. Lang}

\affiliation{Institut f\"{u}r Materialphysik der Universit\"{a}t Wien, Boltzmanngasse
5, A-1090 Wien, Austria}

\begin{abstract}
The non-Ohmic effect of a high electric field on the out-of-plane
magneto-conductivity of a layered superconductor near the superconducting
transition is studied in the frame of the Langevin approach to the
time-dependent Ginzburg-Landau equation. The transverse fluctuation
conductivity is computed in the self-consistent Hartree approximation
for an arbitrarily strong electric field and a magnetic field perpendicular
to the layers. Our results indicate that high electric fields can
be effectively used to suppress the out-of-plane fluctuation conductivity
in high-temperature superconductors and a significant broadening of
the transition induced by a strong electric field is predicted. Extensions
of the results are provided for the case when the electric field is
applied at an arbitrary angle with respect to the layers, as well
as for the three-dimensional anisotropic regime of a strong interlayer
coupling.
\end{abstract}


\maketitle

\section{Introduction}

Outside the critical region above $T_{c}$, in the absence of magnetic
field and for small electric fields, the excess conductivity due to
fluctuations of the superconducting order parameter can be explained
by the Aslamazov-Larkin \cite{Aslamazov68} theory, subsequently extended
by Lawrence and Doniach \cite{Lawrence71} for two-dimensional layered
superconductors, a situation very much resembling the crystal structure
in high-temperature superconductors (HTSC). In the presence of a magnetic
field, the fluctuation transport properties of superconductors were
initially treated \cite{Maki69,Klemm74,FET,Dorin93} in the non-interacting
(Gaussian) fluctuation approach, which predicted a divergence at $T_{c}(H)$
that is, however, not observed. The physical reason is the motion
of vortices providing dissipation and hence a finite flux-flow conductivity.
Ikeda \emph{et al.} \cite{Ikeda91a} and Ullah and Dorsey \cite{UD}
showed that the theoretical divergence can be eliminated by using
the Hartree approximation, which treats self-consistently the quartic
term in the Ginzburg-Landau (GL) free-energy expansion. This approach
was applied for the longitudinal \cite{Ikeda91a,UD} and Hall conductivity,\cite{UD,NE}
as well as for the out-of-plane conductivity,\cite{Livanov97} in
the linear-response approximation for a layered superconductor under
perpendicular magnetic field.

The fluctuation transport properties can be calculated in the linear-response
approximation only for sufficiently weak electric fields that do not
perturb the fluctuation spectrum.\cite{Hurault69} At reasonably high
values of the electric field, the accelaration of the paired electrons
is so large, that on a distance of the order of the coherence length
they change their energy by a value corresponding to the fluctuation
Cooper pair binding energy.\cite{Larkin05} This results in an additional,
field dependent, decay mechanism, and leads to deviation of the current-voltage
characteristics from the Ohm's law. In connection with the low-temperature
superconductors, the non-Ohmic fluctuation conductivity in the absence
of magnetic field has been studied theoretically for the isotropic
case \cite{Schmid69,Tsuzuki70} and found experimentally on thin aluminum
films.\cite{Thomas71,Kajimura71} For a layered superconductor the
issue has been more recently addressed for the in-plane conductivity,
starting from a microscopic approach \cite{Varlamov92} and subsequently
in the frame of the time-dependent Ginzburg-Landau (TDGL) theory,
in the Gaussian \cite{Mishonov02} as well as in the self-consistent
Hartree approximation.\cite{PuicaLangE} Several experimental investigations
of the fluctuation suppression effect of high electric fields in HTSC
were performed for the in-plane paraconductivity in zero magnetic
field,\cite{Soret93,Gorlova95,Kunchur95,Fruchter04,pssc05} and a
good agreement with the theoretical models \cite{Varlamov92,PuicaLangE}
was proven.

The non-linear effect of a strong electric field under the simultaneous
application of a perpendicular magnetic field on the in-plane fluctuation
conductivity and Hall effect was recently addressed by the authors
of the present paper,\cite{PuicaLangM,PuicaLangH} for a layered superconductor
in the Hartree approximation of the TDGL theory. It has been revealed
that the simultaneous application of the two fields results in a slightly
stronger suppression of the superconducting fluctuation conductivity,
compared to the case when the fields are applied individually, while
the relative suppression of the excess Hall conductivity turns out
to be stronger than for the longitudinal one. Experimental investigations
of the fluctuation suppression effect of strong electric fields under
simultaneous application of a magnetic field are however lacking so
far.

Also the \emph{out-of-plane} conductance of a layered superconductor
in the non-Ohmic regime of high electric fields has been, to our knowledge,
neither experimentally nor theoretically investigated up to the present.
The purpose of this paper is to provide a theoretical approach to
this issue, in the frame of the TDGL equation solved in the Hartree
approximation, when both the magnetic and the electric field are applied
perpendicular to the layers. We shall thus be able to find the expression
of the Aslamazov-Larkin (AL) contribution to the out-of-plane fluctuation
conductivity at any values of the magnetic and electric fields, and
predict for the AL term a significant supplementary suppression induced
by a strong electric field in layered HTSC. Based on the proportionality
between the Cooper pair concentration and the density-of-states (DOS)
part of the out-of-plane conductivity, known from the microscopic
theory in the linear response approximation,\cite{Dorin93,Larkin05}
we shall also give an estimate of the non-Ohmic effect on the DOS
contribution. The work will be completed by extending the calculation
of the non-Ohmic fluctuation conductivity to the case when the electric
field is applied at an arbitrary angle with respect to the layers.
This calculation, particularly useful for investigations on vicinal
thin films,\cite{Pedarnig02} is necessary because the non-Ohmic current
produced by a tilted electric field cannot be calculated as the superposition
of the non-Ohmic currents produced separately by the in-plane and
out-of-plane field components, as it was the case in the linear response
approximation. Eventually, the results of the paper will be extended
also to the three-dimensional (3D) anisotropic regime of a strong
interlayer coupling.

\section{\label{Solution}Non-Ohmic out-of-plane fluctuation conductivity}

In order to calculate the non-Ohmic out-of-plane conductivity under
the presence of a magnetic field perpendicular to the layers, we shall
adopt the Langevin approach to the gauge-invariant relaxational TDGL
equation \cite{Schmid69,UD} governing the critical dynamics of the
superconducting order parameter in the $l$-th superconducting plane:\begin{align}
\Gamma_{0}^{-1}\left(\frac{\partial}{\partial t}-i\frac{e_{0}sl}{\hbar}E\right)\psi_{l}+a\psi_{l}+b\left|\psi_{l}\right|^{2}\psi_{l}\label{EQini}\\
-\frac{\hbar^{2}}{2m}\left[\partial_{x}^{2}+\left(\partial_{y}-\frac{ie_{0}}{\hbar}xB\right)^{2}\right]\psi_{l}\nonumber \\
+\frac{\hbar^{2}}{2m_{c}s^{2}}\left(2\psi_{l}-\psi_{l+1}-\psi_{l-1}\right) & =\zeta_{l}\left(x,y,t\right)\;.\nonumber \end{align}
 where $m$ and $m_{c}$ are effective Cooper pair masses in the $ab$-plane
and along the $c$-axis, $s$ is the distance between superconducting
planes, and the pair electric charge is $e_{0}=-2e$. The order parameter
$\psi_{l}$ has the same physical dimension as in the three-dimensional
case, and SI units are used. The GL potential $a=a_{0}\varepsilon$
is parameterized by $a_{0}=\hbar^{2}/2m\xi_{0}^{2}=\hbar^{2}/2m_{c}\xi_{0c}^{2}$
and $\varepsilon=\ln\left(T/T_{0}\right)$, with $T_{0}$ being the
mean-field transition temperature, while $\xi_{0}$ and $\xi_{0c}$
are, respectively, the in-plane and out-of-plane coherence lengths
extrapolated to $T=0$. The order parameter relaxation time $\Gamma_{0}^{-1}$
is given by \cite{Masker69,Cyrot73} $\Gamma_{0}^{-1}=\pi\hbar^{3}/16m\xi_{0}^{2}k_{B}T$.
The magnetic field $\mathbf{B}$, perpendicular to the layers, is
generated by the vector potential in the Landau gauge $\mathbf{A}=\left(0,xB,0\right)$,
with $x$ and $y$ the in-plane coordinates. Since we are interested
in the out-of-plane conductivity, we consider the electric field $\mathbf{E}$
as being applied along the $z$-axis, and generated by the scalar
potential $\varphi_{l}=-Esl$. The Langevin white-noise forces $\zeta_{l}\left(x,y,t\right)$
that describe the thermodynamical fluctuations must satisfy the fluctuation-dissipation
theorem, $\left\langle \zeta_{l}\left(x,y,t\right)\zeta_{l'}^{*}\left(x',y',t'\right)\right\rangle =2\Gamma_{0}^{-1}k_{B}T\delta(x-x')\delta(y-y')\delta(t-t')\delta_{ll'}/s$.

The quartic term in the thermodynamical potential will be treated
in the Hartree approximation,\cite{UD,Penev0} which results in a
linear problem with a modified (renormalized) reduced temperature\begin{equation}
\widetilde{\varepsilon}=\varepsilon+b\left\langle \left|\psi_{l}\right|^{2}\right\rangle /a_{0}\,.\label{RenormEps}\end{equation}
 The out-of-plane fluctuation conductivity $\Delta\sigma_{zz}$ will
be eventually found by calculating the Josephson current density between
the $l$-th and $l+1$-th layers, which in the chosen gauge writes:\begin{eqnarray}
\left\langle \Delta j_{z}^{(l)}\right\rangle  & = & -\frac{i\hbar e_{0}}{2m_{c}s}\left[\left\langle \psi_{l}^{*}\psi_{l+1}\right\rangle -\left\langle \psi_{l}\psi_{l+1}^{*}\right\rangle \right]\,,\label{TransCurrent}\end{eqnarray}
 and further\begin{equation}
\Delta\sigma_{zz}=\left\langle \Delta j_{z}^{(l)}\right\rangle /E\,.\label{szzdef}\end{equation}

It is worth mentioning, at this point, that a slight modification
of the GL free-energy functional for the layered superconductors was
recently proposed,\cite{Ramallo04} which, besides the BCS-type Josephson
coupling, allows for additional interlayer interactions that can contribute
to the condensation energy and give rise to energy savings that enhance
$T_{0}$. The fluctuation spectrum of the proposed functional and
consequently the fluctuation-induced observables like in-plane paraconductivity
and magneto-conductivity were found to be the same as for the usual
Lawrence-Doniach free-energy, once the transition temperature of each
bare layer is renormalized to its value enhanced through the interlayer
energy savings.\cite{Ramallo04} It can be easily verified that this
fact is valid also for the out-of-plane fluctuation conductivity,
since the additional terms in the GL functional of Ref. \onlinecite{Ramallo04}
do not involve the phase of the superconducting parameter, and consequently
do not influence the definition (\ref{TransCurrent}) of the transversal
current density. The results of the present paper will be thus applicable
also for the kind of interlayer coupling proposed in Ref. \onlinecite{Ramallo04}.

We proceed further by introducing the Fourier transform with respect
to the in-plane coordinate $y$, the layer index $l$, and also the
Landau level (LL) representation with respect to the $x$-dependence,
through the relation:\begin{eqnarray}
\psi_{l}(x,y,t) & = & \int\frac{dk}{2\pi}\int_{-\pi/s}^{\pi/s}\frac{dq}{2\pi}\sum_{n\geq0}\psi_{q}(n,k,t)\nonumber \\
 &  & \cdot\mathrm{e}^{-iky}\mathrm{e}^{-iqls}u_{n}\left(x-\frac{\hbar k}{2eB}\right)\,,\label{FourierC}\end{eqnarray}
 where the functions $u_{n}\left(x\right)$ with $n\in\mathbb{N}$
build the orthonormal eigenfunction system of the harmonic oscillator
hamiltonian, so that $\left(-\hbar^{2}\partial_{x}^{2}+4e^{2}B^{2}x^{2}\right)u_{n}\left(x\right)=2\hbar eB\left(2n+1\right)u_{n}\left(x\right)$.
The TDGL equation (\ref{EQini}) will write in the new variables:\begin{eqnarray}
\left[\Gamma_{0}^{-1}\frac{\partial}{\partial t}+\chi\frac{\partial}{\partial q}+a_{0}\widetilde{\varepsilon}_{n}\right.\label{GLEqC}\\
\left.+a_{0}\frac{r}{2}\left(1-\cos qs\right)\right]\psi_{n}(k,q,t) & = & \zeta_{n}(k,q,t)\,,\nonumber \end{eqnarray}
 where the new noise terms $\zeta_{n}(k,q,t)$ are delta-correlated
as $\left\langle \zeta_{n}(k,q,t)\zeta_{n'}^{*}\left(k',q',t'\right)\right\rangle =2\Gamma_{0}^{-1}k_{B}T(2\pi)^{2}\cdot\delta(k-k')\delta(q-q')\delta(t-t')\delta_{nn'}$.
We have also introduced the notations:\begin{eqnarray}
\widetilde{\varepsilon}_{n}=\widetilde{\varepsilon}+\left(2n+1\right)h; & \quad & h=\frac{B}{B_{c2}(0)}=\frac{2e\xi_{0}^{2}B}{\hbar}\,,\nonumber \\
r=\left(\frac{2\xi_{0c}}{s}\right)^{2}; & \quad & \chi=\frac{2eE}{\hbar\Gamma_{0}}\,,\label{Notations}\end{eqnarray}
 with $h$ denoting the reduced magnetic field and $r$ the anisotropy
parameter.

Equation (\ref{GLEqC}) can be solved with the aid of the Green function
technique and has the solution:\begin{eqnarray}
\psi_{n}(k,q,t) & = & \Gamma_{0}\int_{0}^{\infty}d\tau\,\zeta_{n}(k,q-\frac{2eE}{\hbar}\tau,t-\tau)\label{psi-nkqt}\\
 &  & \cdot\exp\left\{ -\Gamma_{0}a_{0}\left[\left(\widetilde{\varepsilon}_{n}+\frac{r}{2}\right)\tau\right.\right.\nonumber \\
 &  & \left.\left.-\frac{r\hbar}{4seE}\left(\sin qs-\sin(q-\frac{2eE}{\hbar}\tau)s\right)\right]\right\} \,,\nonumber \end{eqnarray}
 so that the correlation function between the order parameter in two
layers $l$ and $l'$ will be given by:\begin{eqnarray}
\left\langle \psi_{l}(\mathbf{x},t)\psi_{l'}^{*}(\mathbf{x},t)\right\rangle  & = & \frac{k_{B}T}{2\pi a_{0}\xi_{0}^{2}}h\int_{0}^{\infty}du\sum_{n=0}^{N_{c}}\int_{-\pi/s}^{\pi/s}\frac{dq}{2\pi}\nonumber \\
 &  & \cdot\mathrm{e}^{-iqs\left(l-l'\right)}\exp\left[-u\left(\widetilde{\varepsilon}_{n}+\frac{r}{2}\right)\right.\nonumber \\
 &  & \left.+\frac{r}{2}\frac{\sin pu}{p}\cos\left(qs-pu\right)\right]\,,\label{Corr-ll}\end{eqnarray}
 where the electric field enters the parameter\begin{equation}
p=\frac{\pi es}{16k_{B}T}\, E=\frac{s\sqrt{3}}{\xi_{0}}\frac{E}{E_{0}}\,,\label{p-E}\end{equation}
 with $E_{0}=16\sqrt{3}k_{B}T\,/\,\pi e\xi_{0}$ being the characteristic
electric field defined as in Refs. \onlinecite{Varlamov92} and \onlinecite{Mishonov02}.

We point out that the sum over the LL in Eq. (\ref{Corr-ll}) must
be cut off at some index $N_{c}$, reflecting the inherent UV divergence
of the Ginzburg-Landau theory. The classical \cite{Schmid69,Penev0}
procedure is to suppress the short wavelength fluctuating modes through
a \emph{momentum} (or, equivalently, \emph{kinetic energy}) \emph{cut-off}
condition, which, in terms of the LL representation writes \cite{UD,Penev0}
$\left(2e\hbar B/m\right)\left(n+\frac{1}{2}\right)\leq ca_{0}=c\hbar^{2}/2m\xi_{0}^{2}$,
with the cut-off parameter $c$ of the order of unity. A \emph{total
energy cut-off} was also recently proposed,\cite{Vidal02} whose physical
meaning was shown to follow from the uncertainty principle, and whose
importance is revealed especially at high reduced-temperatures and
magnetic fields close to $B_{c2}(0)$.\cite{Soto04} Formally, the
\emph{}total energy cut-off \emph{}can be obtained from the momentum
cut-off by replacing $c$ with $c-\varepsilon$. However, for low
magnetic fields with respect to $B_{c2}(0)$ and in the critical fluctuation
region, the two cut-off conditions almost coincide quantitatively,
so that we shall apply for simplicity the momentum cut-off procedure.
In terms of the reduced magnetic field $h$, it writes thus $h\left(N_{c}+\frac{1}{2}\right)=c/2$.

Now we are able to apply the expression (\ref{Corr-ll}) in Eqs. (\ref{RenormEps})
and (\ref{TransCurrent}) in order to write, respectively, the self-consistent
Hartree equation and the out-of-plane current density. Thus, after
performing the $q$ integral in the correlation function (\ref{Corr-ll})
taken for $l=l'$, one obtains the renormalizing equation for the
reduced temperature $\widetilde{\varepsilon}$,\begin{equation}
\widetilde{\varepsilon}=\ln\frac{T}{T_{0}}+2gTh\sum_{n=0}^{N_{c}}\int_{0}^{\infty}du\,\mathrm{e}^{-u\left(\widetilde{\varepsilon}_{n}+\frac{r}{2}\right)}I_{0}\left(\frac{r}{2p}\sin(pu)\right)\,,\label{self-consist-C}\end{equation}
 where $I_{0}(x)$ is the modified Bessel function and the parameter\begin{equation}
g=\frac{2\mu_{0}\kappa_{\mathrm{GL}}^{2}e^{2}\xi_{0}^{2}k_{B}}{\pi\hbar^{2}s}\label{g}\end{equation}
 was introduced according to the expression of the quartic term coefficient
\cite{UD} $b=\mu_{0}\kappa_{\mathrm{GL}}^{2}e_{0}^{2}\hbar^{2}/2m^{2}$,
with $\kappa_{\mathrm{GL}}$ being the in-plane Ginzburg-Landau parameter
$\kappa_{\mathrm{GL}}=\lambda_{0}/\xi_{0}$.

In an analogous manner, after computing the correlation function (\ref{Corr-ll})
for $l'=l+1$, and using the current density definition (\ref{TransCurrent}),
one can eventually obtain the out-of-plane fluctuation conductivity
under arbitrary magnetic and electric fields,\begin{widetext}\begin{eqnarray}
\Delta\sigma_{zz}^{\mathrm{AL}}(E,B) & = & \frac{e^{2}s\, r\, h}{32\hbar\xi_{0}^{2}}\sum_{n=0}^{N_{c}}\int_{0}^{\infty}du\,\mathrm{e}^{-u\left(\widetilde{\varepsilon}_{n}+\frac{r}{2}\right)}\frac{\sin(pu)}{p}\, I_{1}\left(\frac{r}{2p}\sin(pu)\right)\,,\label{sigmazz-C}\end{eqnarray}
 with $I_{1}(x)$ the modified Bessel function of first order. The
integrals in Eqs. (\ref{self-consist-C}) and (\ref{sigmazz-C}) are
convergent provided $\widetilde{\varepsilon}+\frac{r}{2}+h>0$, so
that $\widetilde{\varepsilon}_{n}+\frac{r}{2}>0$ for any LL index
$n$. This condition is however assured while solving Eq. (\ref{self-consist-C})
for the parameter $\widetilde{\varepsilon}$ at any temperature $T$.

It is useful to write Eq. (\ref{sigmazz-C}) also when the cut-off
is neglected, i.e. for $c,\, N_{c}\rightarrow\infty$:\begin{eqnarray}
\Delta\sigma_{zz}^{\mathrm{AL-NoCut}}(E,B) & = & \frac{e^{2}s\, r}{64\hbar\xi_{0}^{2}}\int_{0}^{\infty}du\,\frac{2hu\,\mathrm{e}^{-uh}}{1-\exp\left(-2hu\right)}\mathrm{e}^{-u\left(\widetilde{\varepsilon}+\frac{r}{2}\right)}\cdot\frac{\sin(pu)}{pu}\, I_{1}\left(\frac{r}{2p}\sin(pu)\right)\,.\label{sigmazzC-noCut}\end{eqnarray}
\end{widetext}

This slightly simpler formula provides a good approximation for Eq.
(\ref{sigmazz-C}) in the temperature region close to the transition
(where $\widetilde{\varepsilon}\ll c$) and for small magnetic fields
(for which $N_{c}=(c-h)/2h$ is already high). The cut-off procedure
remains however essential for calculating the Cooper pairs density
$\left\langle \left|\psi_{l}(\mathbf{x},t)\right|^{2}\right\rangle $
contained in the renormalization equation (\ref{self-consist-C}),
since the $u$-integral would be divergent for $c,\, N_{c}\rightarrow\infty$.
As we shall see in Section \ref{Non-layered}, Eq. (\ref{sigmazzC-noCut})
can be also directly transformed in order to find the three dimensional
limit (i.e. for $s\rightarrow0$) of the non-Ohmic fluctuation conductivity
$\left.\Delta\sigma_{zz}^{\mathrm{AL}}\right|_{\mathrm{NoCut}}^{(3\mathrm{D})}$
parallel to the magnetic field, when the cut-off is neglected.

In Eq. (\ref{sigmazz-C}) we have explicitly specified that the out-of-plane
fluctuation conductivity $\Delta\sigma_{zz}^{\mathrm{AL}}$ corresponds
to the Aslamazov-Larkin (AL) fluctuation process, since the phenomenological
Ginzburg-Landau theory cannot account for indirect contributions like
the density-of-states (DOS) and Maki-Thompson (MT) terms, which can
be found only from a microscopical approach. However, whereas the
DOS and MT contributions to the in-plane paraconductivity and Hall
effect coefficient are known to be negligible near the superconducting
transition with respect to the AL term due to the more singular behavior
of the latter, an investigation of the out-of-plane conductivity needs
taking into account also the DOS term, which can compete with the
AL one especially for highly anisotropic materials, as pointed out
by Larkin and Varlamov.\cite{Larkin05} The reason is that the DOS
contribution to the out-of-plane fluctuation conductivity turns out
to be proportional to a lower order of the interlayer transparency
than the AL one, as shown by the microscopical approach in the linear
response approximation.\cite{Dorin93,Larkin05} We shall tentatively
give in Section \ref{DOSestimation} an estimate of the DOS contribution
to $\Delta\sigma_{zz}$ also for an arbitrarily strong electric field,
after discussing the limit values of expressions (\ref{sigmazz-C})
and (\ref{self-consist-C}) in the cases of a vanishing magnetic or
electric field.

\section{\label{LimitCases}Limit cases $B\rightarrow0$ and/or $E\rightarrow0$}

The Ohmic out-of-plane fluctuation conductivity in the presence of
a magnetic field but in an infinitesimally small electric field can
be easily obtained by taking the limit $p\rightarrow0$ in Eq. (\ref{sigmazz-C})
, which acquires thus the form:\begin{eqnarray}
\left.\Delta\sigma_{zz}^{\mathrm{AL}}(B)\right|_{E\rightarrow0} & = & \frac{e^{2}s\, r^{2}}{64\hbar\xi_{0}^{2}}h\sum_{n=0}^{N_{c}}\left[\widetilde{\varepsilon}_{n}\left(\widetilde{\varepsilon}_{n}+r\right)\right]^{-\frac{3}{2}},\label{sigmazzE0}\end{eqnarray}
 The expression (\ref{sigmazzE0}) matches thus the result previously
obtained within the diagrammatic microscopic approach, for Gaussian
fluctuations (i.e. with $\widetilde{\varepsilon}=\varepsilon$), in
the linear response approximation.\cite{Dorin93} Analogously, the
self-consistent equation (\ref{self-consist-C}) becomes, in the same
limit,\begin{eqnarray}
\left.\widetilde{\varepsilon}\right|_{E\rightarrow0,B>0} & = & \ln\frac{T}{T_{0}}+2gTh\sum_{n=0}^{N_{c}}\left[\widetilde{\varepsilon}_{n}\left(\widetilde{\varepsilon}_{n}+r\right)\right]^{-\frac{1}{2}},\label{SelfConstE0}\end{eqnarray}
 which represents the Hartree renormalization equation in the linear
response approximation under an applied magnetic field found in Ref.
\onlinecite{UD}.

The other limit case, namely for vanishing magnetic field but under
a finite applied electric field, needs taking the limit $h\rightarrow0$
in Eqs. (\ref{sigmazz-C}) and (\ref{self-consist-C}) after performing
the sum over Landau levels and taking into account the cut-off condition,
so that the Hartree self-consistent renormalization (\ref{self-consist-C})
and the out-of-plane fluctuation conductivity (\ref{sigmazz-C}) become,
respectively,\begin{widetext}\begin{eqnarray}
\left.\widetilde{\varepsilon}\right|_{E>0,B=0} & = & \ln\frac{T}{T_{0}}+gT\int_{0}^{\infty}du\,\frac{1-\mathrm{e}^{-cu}}{u}\mathrm{e}^{-u\left(\widetilde{\varepsilon}+\frac{r}{2}\right)}I_{0}\left(\frac{r}{2p}\sin(pu)\right)\,,\label{selconsistCB0}\\
\left.\Delta\sigma_{zz}^{\mathrm{AL}}\right|_{E>0,B=0} & = & \frac{e^{2}s\, r}{64\hbar\xi_{0}^{2}}\int_{0}^{\infty}du\,\left(1-\mathrm{e}^{-cu}\right)\mathrm{e}^{-u\left(\widetilde{\varepsilon}+\frac{r}{2}\right)}\cdot\frac{\sin(pu)}{pu}\, I_{1}\left(\frac{r}{2p}\sin(pu)\right)\,.\label{sigmazzB0}\end{eqnarray}
\end{widetext}

It can be easily verified, by using the integral identities of the
Bessel functions $I_{0}$ and $I_{1}$, that in the further limit
$E\rightarrow0$, Eq. (\ref{selconsistCB0}) becomes\begin{equation}
\left.\widetilde{\varepsilon}\right|_{E=0,B=0}=\ln\frac{T}{T_{0}}+2gT\,\ln\frac{\sqrt{\widetilde{\varepsilon}+c}+\sqrt{\widetilde{\varepsilon}+c+r}}{\sqrt{\widetilde{\varepsilon}}+\sqrt{\widetilde{\varepsilon}+r}}\,,\label{SelfConstEB0}\end{equation}
 as also found in Ref. \onlinecite{Penev0}, while expression (\ref{sigmazzB0})
takes, respectively, the form:\begin{eqnarray}
\left.\Delta\sigma_{zz}^{\mathrm{AL}}\right|_{E=0,B=0} & = & \frac{e^{2}s}{32\hbar\xi_{0}^{2}}\left[\frac{\widetilde{\varepsilon}+\frac{r}{2}}{\sqrt{\widetilde{\varepsilon}\left(\widetilde{\varepsilon}+r\right)}}\right.\label{sigmazzE0B0}\\
 &  & \left.-\frac{\widetilde{\varepsilon}+c+\frac{r}{2}}{\sqrt{\left(\widetilde{\varepsilon}+c\right)\left(\widetilde{\varepsilon}+c+r\right)}}\right]\,.\nonumber \end{eqnarray}
 If one neglects the cut-off procedure (i.e. $c\rightarrow\infty$),
expression (\ref{sigmazzE0B0}) matches the result obtained in Ref.
\onlinecite{Larkin05} for the AL out-of-plane fluctuation conductivity
in the linear response limit and in the absence of magnetic field,
with the difference that in Ref. \onlinecite{Larkin05}, based on
the Gaussian approximation, the reduced temperature $\varepsilon=\ln\left(T/T_{0}\right)$
is present instead of our Hartree renormalized $\widetilde{\varepsilon}$.

The Hartree renormalization procedure consists in using the reduced
temperature parameter $\widetilde{\varepsilon}$ instead of $\varepsilon=\ln\left(T/T_{0}\right)$,
by solving Eq. (\ref{SelfConstEB0}). This procedure causes the critical
temperature to shift downwards with respect to the bare mean-field
transition temperature $T_{0}$. In analogy with the Gaussian fluctuation
case, we shall adopt as definition for the critical temperature the
vanishing of the reduced temperature, $\widetilde{\varepsilon}=0$,
where the fluctuation conductivity, given by Eq. (\ref{sigmazzE0B0}),
diverges. In practice, one knows experimentally the actual critical
temperature $T_{c0}$ measured at very low electrical field and with
zero magnetic field, so that the relationship between $T_{c0}$ and
$T_{0}$ will be found by putting $\widetilde{\varepsilon}=0$ in
Eq. (\ref{SelfConstEB0}). It writes :\cite{PuicaLangE}\begin{equation}
T_{0}=T_{c0}\left[\sqrt{c/r}+\sqrt{1+(c/r)}\right]^{2gT_{c0}}\,.\label{T0-Tc0}\end{equation}
 Now, having $T_{0}$ one can use Eq. (\ref{self-consist-C}) for
any temperature $T$ and fields $E$ and $B$ in order to find the
actual renormalized $\widetilde{\varepsilon}(T,E,B)$.

\section{\label{DOSestimation}Estimation of the non-Ohmic DOS contribution}

In the microscopic approach of Ref. \onlinecite{Dorin93}, valid in
the linear response approximation (i.e. for vanishing electric field)
and for non-interacting (Gaussian) fluctuations, the DOS contribution
to the out-of-plane fluctuation conductivity under a magnetic field
is found to amount, in our notations:\begin{eqnarray}
\left.\Delta\sigma_{zz}^{\mathrm{DOS}}\right|_{E\rightarrow0,B>0}^{\mathrm{Gaussian}} & = & -\frac{e^{2}s\,\kappa r}{8\hbar\xi_{0}^{2}}\, h\sum_{n=0}^{N_{c}}\left[\widetilde{\varepsilon}_{n}\left(\widetilde{\varepsilon}_{n}+r\right)\right]^{-\frac{1}{2}}\,,\label{DorinDOS}\end{eqnarray}
 where the parameter $\kappa$ depends on the impurity scattering
time $\tau$ and temperature:\cite{Dorin93}\begin{equation}
\kappa=\frac{1}{\pi^{2}}\cdot\frac{-\psi'\left(\frac{1}{2}+\frac{1}{4\pi}\frac{\hbar}{\tau k_{B}T}\right)+\frac{1}{2\pi}\frac{\hbar}{\tau k_{B}T}\psi''\left(\frac{1}{2}\right)}{\psi\left(\frac{1}{2}+\frac{1}{4\pi}\frac{\hbar}{\tau k_{B}T}\right)-\psi\left(\frac{1}{2}\right)-\frac{1}{4\pi}\frac{\hbar}{\tau k_{B}T}\psi'\left(\frac{1}{2}\right)}\,,\label{ka}\end{equation}
 with $\psi\left(x\right)$ the Euler digamma function. One can notice
that Eq. (\ref{DorinDOS}) can be written as:\begin{equation}
\left.\Delta\sigma_{zz}^{\mathrm{DOS}}\right|_{E\rightarrow0,B>0}^{\mathrm{Gaussian}}=-\frac{e^{2}\pi\hbar\,\kappa}{2m_{c}k_{B}T}\left.\left\langle \left|\psi\right|^{2}\right\rangle \right|_{E\rightarrow0,B>0}^{\mathrm{Gaussian}}\,,\label{proportDOS}\end{equation}
 where\begin{eqnarray}
\left.\left\langle \left|\psi\right|^{2}\right\rangle \right|_{E\rightarrow0,B>0}^{\mathrm{Gaussian}} & = & \frac{mk_{B}T}{\pi\hbar^{2}s}\, h\sum_{n=0}^{N_{c}}\left[\widetilde{\varepsilon}_{n}\left(\widetilde{\varepsilon}_{n}+r\right)\right]^{-\frac{1}{2}}\label{PairDensity}\end{eqnarray}
 is the Coooper pairs density for vanishing electric field and in
the Gaussian approximation (i.e. with $\widetilde{\varepsilon}=\varepsilon$),
as one can infer from the general correlation function (\ref{Corr-ll})
in the $E\rightarrow0$ limit. The proportionality between the DOS
fluctuation conductivity and the Cooper pair concentration in Eq.
(\ref{proportDOS}) is qualitatively easy to be grasped, since the
DOS contribution means in fact the reduction of the normal state conductivity
due to the decrease of the one-electron density of states, which reduction
is in turn proportional to the superfluid density.\cite{Larkin05}

We may assume that the proportionality (\ref{proportDOS}) will hold
also in the case of an arbitrarily strong electric field and in the
Hartree approximation, so that the DOS contribution to the out-of-plane
fluctuation conductivity can be generally written:\begin{eqnarray}
\Delta\sigma_{zz}^{\mathrm{DOS}}\left(E,B\right) & = & -\frac{e^{2}s\,\kappa r}{8\hbar\xi_{0}^{2}}\, h\sum_{n=0}^{N_{c}}\label{sigmazzDOS}\\
 &  & \cdot\int_{0}^{\infty}du\,\mathrm{e}^{-u\left(\widetilde{\varepsilon}_{n}+\frac{r}{2}\right)}I_{0}\left(\frac{r}{2p}\sin(pu)\right)\,,\nonumber \end{eqnarray}
 where Eqs. (\ref{RenormEps}) and (\ref{self-consist-C}) are to
be compared in order to reveal the Cooper pair density $\left\langle \left|\psi\right|^{2}\right\rangle $.
Analogous with Eq. (\ref{sigmazzB0}) we can consequently write the
DOS contribution also in the vanishing magnetic field limit,\begin{eqnarray}
\left.\Delta\sigma_{zz}^{\mathrm{DOS}}\right|_{E>0,B=0} & = & -\frac{e^{2}s\,\kappa r}{16\hbar\xi_{0}^{2}}\int_{0}^{\infty}du\,\frac{1-\mathrm{e}^{-cu}}{u}\label{sigmazzDOSB0}\\
 &  & \cdot\mathrm{e}^{-u\left(\widetilde{\varepsilon}+\frac{r}{2}\right)}I_{0}\left(\frac{r}{2p}\sin(pu)\right)\,.\nonumber \end{eqnarray}
 The equations (\ref{sigmazzDOS}) and (\ref{sigmazzDOSB0}) remain
however to be confirmed or refuted by a microscopic approach.\\

\section{\label{Illustration}Example: optimally doped $\textrm{YBa}{}_{2}\textrm{Cu}{}_{3}\textrm{O}{}_{6+x}$}

In order to illustrate the main features of our model, we take as
example a common HTSC material, like the optimally doped YBa$_{2}$Cu$_{3}$O$_{6+x}$
(YBCO). Typical characteristic parameters are then: $s=1.17$ nm for
the interlayer distance, $\xi_{0}=1.2$ nm and $\xi_{0c}=0.14$ nm
for the zero-temperature-extrapolated in-plane and out-of-plane coherence
lengths, respectively, $\kappa_{\mathrm{GL}}=70$ for the Ginzburg-Landau
parameter, $T_{c0}=92$ K for the critical temperature under very
small electric and zero magnetic field, and the parameter $\kappa=3.57$
that corresponds to a scattering time $\tau\simeq30\,\mathrm{fs}$
in Eq. (\ref{ka}). It must be stated that the form of the normal-state
background chosen for the temperature region masked by the onset of
the superconductivity can be crucial for an eventual comparison with
the experiment. There is however no consensus whether the peculiarities
of the out-of-plane resistivity, namely its peak and its non-metallic
character just above $T_{c0}$, as observed in the oxygen-deficient
YBCO and the more anisotropic Bi$_{2}$Sr$_{2}$CaCu$_{2}$O$_{8+x}$,
are mainly due to the competition between the fluctuation AL and DOS
contributions, as illustrated in Ref. \onlinecite{Livanov97} by succesfull
fits at different magnetic fields in the Ohmic regime while assuming
a metallic linear extrapolation for the normal-state resistivity,
or to the inherent behavior of the normal-state itself, as suggested
by analysis of conductivity in incoherent layered crystals.\cite{Levin04}
Since in this paper we focus on the fluctuation conductivity in the
non-Ohmic regime and need the normal-state background only for illustration
purposes of the resistivity characteristics, we shall further assume,
for simplicity, an out-of-plane normal state resistivity almost constant
near the transition, with a typical value $\rho_{c}^{\mathrm{N}}=4\,\mathrm{m\Omega cm}$
for optimally doped YBCO.\cite{Heine99}

\begin{figure*}
\includegraphics[  width=12.5cm]{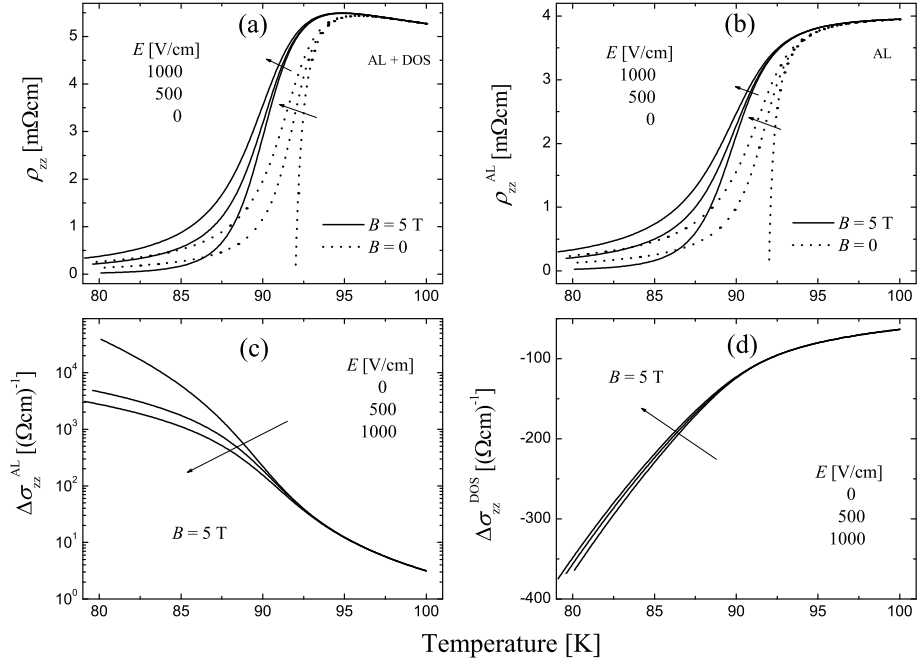}

\caption{(a) Out-of-plane resistivity as a function of temperature in YBa$_{2}$Cu$_{3}$O$_{6+x}$,
for two values of the magnetic field, at several magnitudes of the
electric field, when both AL and DOS fluctuation contributions are
taken into account. The material parameters are given in the text.
The arrows show the increasing electric field direction. (b) Same
as (a), when only the AL part is considered. (c) The out-of-plane
AL conductivity, at different magnitudes of the electric field, at
$B=5\,\mathrm{T}$. (d) The DOS contribution to the out-of-plane conductivity,
at different electric fields and fixed magnetic field.\label{FigRhozz}}
\end{figure*}

In Fig. \ref{FigRhozz}a the out-of-plane resistivity is presented,
when both AL and DOS contributions are taken into account, while Fig.
\ref{FigRhozz}b shows the effect of the AL term alone, if the same
normal state background is assumed. One can notice the supplementary
broadening of the transition induced by a strong electric field, and
also the relative reduction of the non-Ohmic effect when a magnetic
field is simultaneously applied. The effect of various electric fields,
at a fixed magnetic field, on the AL and DOS fluctuation conductivities
is detailed in Figs. \ref{FigRhozz}c and \ref{FigRhozz}d, respectively.
It turns out that the non-Ohmic effect is important only for the AL
term, while the DOS one is little affected by an electric field of
experimentally accessible strength. This behavior stems from the peculiar
dependence of Eq. (\ref{sigmazzDOS}) on the electric field only in
the argument of the Bessel function.

It is worth mentioning that for Bi$_{2}$Sr$_{2}$CaCu$_{2}$O$_{8+x}$,
for which the DOS contribution as such competes stronger with the
AL one due to the higher anisotropy, the estimate of the non-Ohmic
effect on the DOS term turns out however to be even more insignificant
than for YBa$_{2}$Cu$_{3}$O$_{6+x}$, and thus almost undiscernable
in the same range of electric fields. The reason is the much smaller
anisotropy parameter $r$ which reduces the effect of the Bessel function
factor in Eq. (\ref{sigmazzDOS}).

\section{\label{TiltedE}Non-Ohmic conduction for a tilted electric field}

In the linear response approximation, the current produced by an arbitrarily
oriented electric field can be simply obtained by superposing the
currents generated separately by its components. However, this is
not anymore the case in the non-Ohmic regime of a strong electric
field. As we shall see below, the current components will depend now
on all the field components, and not only to the particular one corresponding
to the respective axis. This case requires therefore a special treatment,
where both the in-plane and out-of-plane electric field components
are to be included from the beginning in the TDGL equation. This calculation
would be particularly useful if the investigation of the out-of-plane
non-Ohmic conduction were performed on vicinal thin films,\cite{Pedarnig02}
where a mixture of the in-plane and out-of-plane transport properties
is assessed, since the injected current has a slanted direction with
respect to the crystallographic axes.

We shall consider in the following the case of an electric field $\mathbf{E}$
applied on a layered superconductor at an angle $\theta$ with the
$c$-axis, having thus the components $\left(E_{x}=E\sin\theta,\, E_{y}=0,\, E_{z}=E\cos\theta\right)$,
and generated by the scalar potential $\varphi=-E_{x}x-E_{z}sl$.
Since the presence of a magnetic field at arbitrary direction would
overcomplicate the calculations, we shall consider in the following
the case of a zero magnetic field. The TDGL equation analogous to
Eq. (\ref{GLEqC}) will write in this case:\begin{eqnarray}
\left[\Gamma_{0}^{-1}\frac{\partial}{\partial t}+\chi\sin\theta\frac{\partial}{\partial k}+\chi\cos\theta\frac{\partial}{\partial q}\right.\label{GLEqTilted}\\
+a_{0}\left(\widetilde{\varepsilon}+\xi_{0}^{2}k^{2}+\xi_{0}^{2}k_{y}^{2}\right)\nonumber \\
\left.+a_{0}\frac{r}{2}\left(1-\cos qs\right)\right]\psi(k,k_{y},q,t) & = & \zeta(k,k_{y},q,t)\,,\nonumber \end{eqnarray}
 where the Fourier transformed order parameter $\psi(k,k_{y},q,t)$
is given by\begin{eqnarray}
\psi_{l}(x,y,t) & = & \int\frac{dk}{2\pi}\int\frac{dk_{y}}{2\pi}\int_{-\pi/s}^{\pi/s}\frac{dq}{2\pi}\label{FourierB0}\\
 &  & \cdot\mathrm{e}^{-ikx}\mathrm{e}^{-ik_{y}y}\mathrm{e}^{-iqls}\psi(k,k_{y},q,t)\,,\nonumber \end{eqnarray}
 and the noise terms are delta-correlated such as$\left\langle \zeta(k,k_{y},q,t)\zeta^{*}\left(k',k_{y}',q',t'\right)\right\rangle =2\Gamma_{0}^{-1}k_{B}T(2\pi)^{3}\cdot\delta(k-k')\delta(k_{y}-k_{y}')\delta(q-q')\delta(t-t')$.
It can be verified that with the aid of the Green function in the
three variables $\left(k,q,t\right)$ for Eq. (\ref{GLEqTilted}),
the solution for the Fourier-transformed order parameter writes:\begin{widetext}

\begin{eqnarray}
\psi(k,k_{y},q,t) & = & \Gamma_{0}\int_{0}^{\infty}d\tau\,\zeta(k-\frac{2eE_{x}}{\hbar}\tau,\, k_{y},\, q-\frac{2eE_{z}}{\hbar}\tau,\, t-\tau)\label{psi-Etilted}\\
 &  & \cdot\exp\left\{ -a_{0}\Gamma_{0}\tau\left[\left(\widetilde{\varepsilon}+\xi_{0}^{2}k_{y}^{2}+\xi_{0}^{2}k^{2}+\frac{r}{2}\right)-\xi_{0}^{2}k\frac{2eE_{x}}{\hbar}\tau+\frac{\xi_{0}^{2}}{3}\left(\frac{2eE_{x}}{\hbar}\right)^{2}\tau^{2}\right]\right\} \nonumber \\
 &  & \cdot\exp\left[\frac{a_{0}\Gamma_{0}\hbar r}{2eE_{z}s}\sin\frac{eE_{z}s\tau}{\hbar}\cos\left(qs-\frac{eE_{z}s\tau}{\hbar}\right)\right]\,.\nonumber \end{eqnarray}
 Consequently, the general order parameter correlation function that
will allow us to calculate the Cooper pair density as well as the
current density along the $x$ and $z$ axes will write:\begin{eqnarray*}
\left\langle \psi_{l}(x,y,t)\psi_{l'}^{*}(x',y,t)\right\rangle  & = & \frac{k_{B}T}{a_{0}}\int_{0}^{\infty}du\left(\int\frac{dk}{2\pi}\int\frac{dk_{y}}{2\pi}\right)_{\xi_{0}^{2}\left(k_{y}^{2}+k^{2}\right)\leq c}\int_{-\pi/s}^{\pi/s}\frac{dq}{2\pi}\\
 &  & \cdot\mathrm{e}^{-ik\left(x-x'\right)}\mathrm{e}^{-iqs\left(l-l'\right)}\exp\left[-u\left(\widetilde{\varepsilon}+\xi_{0}^{2}k_{y}^{2}+\xi_{0}^{2}k^{2}+\frac{r}{2}\right)-4\left(\frac{E\sin\theta}{E_{0}}\right)^{2}u^{3}\right]\\
 &  & \cdot\exp\left(2\sqrt{3}\xi_{0}k\frac{E\sin\theta}{E_{0}}u^{2}\right)\exp\left[\frac{r}{2}\frac{\sin\left(pu\cos\theta\right)}{p\cos\theta}\cos\left(qs-pu\cos\theta\right)\right]\,,\end{eqnarray*}
 where $E_{0}$ and $p$ are defined in Eq. (\ref{p-E}). One should
also notice the cut-off condition for the in-plane momentum $\xi_{0}^{2}\left(k_{y}^{2}+k^{2}\right)\leq c$.

We are now able to summarize the results obtained for the Hartree
renormalization equation and for the current density components as
being, respectively,\begin{eqnarray}
\widetilde{\varepsilon}_{\left(E_{x},E_{z}\right)} & = & \ln\frac{T}{T_{0}}+gT\int_{0}^{\infty}du\,\mathrm{e}^{-u\left(\widetilde{\varepsilon}+\frac{r}{2}\right)-4\left(\frac{E_{x}}{E_{0}}\right)^{2}u^{3}}I_{0}\left(\frac{r}{2}\frac{\sin(p_{z}u)}{p_{z}}\right)\int_{0}^{c}dw\,\mathrm{e}^{-uw}I_{0}\left(2\sqrt{3}\sqrt{w}\frac{E_{x}}{E_{0}}u^{2}\right)\,,\label{epstilde-Etilted}\\
\Delta j_{x}\left(E_{x},E_{z}\right) & = & \frac{e^{2}}{16\hbar s}E_{x}\int_{0}^{\infty}du\, u^{2}\mathrm{e}^{-u\left(\widetilde{\varepsilon}+\frac{r}{2}\right)-4\left(\frac{E_{x}}{E_{0}}\right)^{2}u^{3}}I_{0}\left(\frac{r}{2}\frac{\sin(p_{z}u)}{p_{z}}\right)\int_{0}^{c}dw\, w\mathrm{e}^{-uw}\left(I_{0}-I_{2}\right)_{\left(2\sqrt{3}\sqrt{w}\frac{E_{x}}{E_{0}}u^{2}\right)}\,,\label{jx-Etilted}\\
\Delta j_{z}\left(E_{x},E_{z}\right) & = & \frac{e^{2}}{16\hbar s}\gamma_{a}^{2}E_{z}\int_{0}^{\infty}du\,\mathrm{e}^{-u\left(\widetilde{\varepsilon}+\frac{r}{2}\right)-4\left(\frac{E_{x}}{E_{0}}\right)^{2}u^{3}}\cdot\frac{\sin(p_{z}u)}{p_{z}}I_{1}\left(\frac{r}{2}\frac{\sin(p_{z}u)}{p_{z}}\right)\int_{0}^{c}dw\,\mathrm{e}^{-uw}I_{0}\left(2\sqrt{3}\sqrt{w}\frac{E_{x}}{E_{0}}u^{2}\right)\,,\label{jz-Etilted}\end{eqnarray}
 where $p_{z}=p\cos\theta=\left(s\sqrt{3}/\xi_{0}\right)\left(E_{z}/E_{0}\right)$,
$\gamma_{a}=\xi_{0c}/\xi_{0}=\sqrt{m/m_{c}}$ is the anisotropy factor,
and $I_{0}$, $I_{1}$ and $I_{2}$ are the modified Bessel functions.
The current density components can be written in a simpler form if
the cut-off procedure is neglected ($c\rightarrow\infty$), namely\begin{eqnarray}
\Delta j_{x}^{\mathrm{NoCut}}\left(E_{x},E_{z}\right) & = & \frac{e^{2}}{16\hbar s}E_{x}\int_{0}^{\infty}du\,\mathrm{e}^{-u\left(\widetilde{\varepsilon}+\frac{r}{2}\right)-\left(E_{x}/E_{0}\right)^{2}u^{3}}I_{0}\left(\frac{r}{2}\frac{\sin(p_{z}u)}{p_{z}}\right)\,,\label{jx-Etilted-NoCut}\\
\Delta j_{z}^{\mathrm{NoCut}}\left(E_{x},E_{z}\right) & = & \frac{e^{2}}{16\hbar s}\gamma_{a}^{2}E_{z}\int_{0}^{\infty}du\,\mathrm{e}^{-u\left(\widetilde{\varepsilon}+\frac{r}{2}\right)-\left(E_{x}/E_{0}\right)^{2}u^{3}}\cdot\frac{\sin(p_{z}u)}{p_{z}u}I_{1}\left(\frac{r}{2}\frac{\sin(p_{z}u)}{p_{z}}\right)\,.\label{jz-Etilted-NoCut}\end{eqnarray}
\end{widetext}

Besides the superconducting fluctuation contribution $\left(\Delta j_{x},\Delta j_{z}\right)$,
also the normal state conduction must be considered in the total current
density, which will be thus given by:\begin{eqnarray}
j_{x} & = & \Delta j_{x}\left(E_{x},E_{z}\right)+\sigma_{ab}^{\mathrm{N}}E_{x}\,;\label{TotalJ}\\
j_{z} & = & \Delta j_{z}\left(E_{x},E_{z}\right)+\sigma_{c}^{\mathrm{N}}E_{z}\,,\nonumber \end{eqnarray}
 if one supposes Ohmic in-plane and out-of-plane normal state conductivities
$\sigma_{ab}^{\mathrm{N}}$ and $\sigma_{c}^{\mathrm{N}}$, respectively.
As it is generally the case for an anisotropic conductor, the current
is not parallel to the field, unless the latter is applied along one
of the principal axes of the material. Moreover, unlike the Ohmic
regime, the current densities components $\Delta j_{x}$ and $\Delta j_{z}$
in Eqs. (\ref{jx-Etilted}) and (\ref{jz-Etilted}) or in Eqs. (\ref{jx-Etilted-NoCut})
and (\ref{jz-Etilted-NoCut}) depend generally on both $E_{x}$ and
$E_{z}$, so that the non-Ohmic effect of an arbitrarily oriented
electric field can not be reduced to the superposition of the non-Ohmic
currents produced separately by the in-plane and out-of-plane field
components.

It must be also mentioned that the superconducting fluctuation current$\left(\Delta j_{x},\Delta j_{z}\right)$
in Eq. (\ref{TotalJ}), as inferred in the GL approach, regards only
the AL fluctuation process.

\section{\label{Non-layered}Results in the 3D limit}

\subsection{General case of non-zero magnetic and electric fields}

The results obtained in the previous sections for a layered superconductor
cannot be transformed for the three-dimensional case by directly imposing
the 3D condition $s\rightarrow0$ (or $r\rightarrow\infty$), because
in the layered model it was assumed that a cut-off in the $z$ direction
is not necessary, since the out-of-plane momentum $q$ was already
confined to the interval $\left[-\pi/s,\pi/s\right]$. However, in
the 3D case, a cut-off condition is necessary for all the three momentum
components, $\left(k_{x},k_{y},q\right)$. Assuming isotropy in the
$xy$-plane, the 3D cut-off condition for the total kinetic energy
will write thus, in the absence of a magnetic field, \begin{equation}
\frac{\hbar^{2}k_{x}^{2}}{2m}+\frac{\hbar^{2}k_{y}^{2}}{2m}+\frac{\hbar^{2}q^{2}}{2m_{c}}\leq ca_{0}\quad\mathrm{or}\quad\left(k_{x}^{2}+k_{y}^{2}\right)\xi_{0}^{2}+q^{2}\xi_{0c}^{2}\leq c\,.\label{Cut-off-3Dk}\end{equation}
 If a magnetic field $B$ is applied along the $z$-axis, the cut-off
condition will be written though in terms of the Landau level $n$,
as:

\begin{equation}
a_{0}h(2n+1)+\frac{\hbar^{2}q^{2}}{2m_{c}}\leq ca_{0}\quad\mathrm{or}\quad h(2n+1)+q^{2}\xi_{0c}^{2}\leq c\,.\label{Cut-off-3D}\end{equation}

The kinetic energy $\hbar^{2}q^{2}/2m_{c}$ will replace therefore
the interlayer coupling energy $a_{0}r\left(1-\cos qs\right)/2$ from
the layered case. This can be formally performed by taking the cosine
function in the small-$q$-limit, such as:\begin{equation}
\frac{r}{2}\left(1-\cos qs\right)\approx\frac{r\, q^{2}s^{2}}{4}=q^{2}\xi_{0c}^{2}\equiv q'^{2}\,,\label{KinEnZ}\end{equation}
 which can be interpreted by saying that the 3D behavior is approached
when the size of the Cooper pairs along the $z$-axis is so large
that the peculiarities of the layered structure do not play a role
any more, meaning that only small values of the out-of-plane momentum
$q$ are important in the integrations. This regime indeed occurs
for the layered superconductors in the near vicinity of the transition
temperature, where the fluctuations acquire anisotropic 3D character.

In the presence of an electric field $E$ along the $z$-axis, one
must solve now the TDGL equation in the form: \begin{equation}
\left[\Gamma_{0}^{-1}\frac{\partial}{\partial t}+\chi\frac{\partial}{\partial q}+a_{0}\widetilde{\varepsilon}_{n}+a_{0}q^{2}\xi_{0c}^{2}\right]\psi_{n}(k,q,t)=\zeta_{n}(k,q,t),\label{TDGL-3D}\end{equation}
 so that the solution is found to be:\begin{widetext}\begin{eqnarray}
\psi_{n}(k,q,t) & = & \Gamma_{0}\int_{0}^{\infty}d\tau\,\zeta_{n}(k,q-\frac{2eE}{\hbar}\tau,t-\tau)\cdot\exp\left\{ -\Gamma_{0}\tau\left[a_{0}\widetilde{\varepsilon}_{n}+\frac{e^{2}E^{2}}{6m_{c}}\tau^{2}+\frac{\hbar^{2}}{2m_{c}}\left(q-\frac{eE}{\hbar}\tau\right)^{2}\right]\right\} \,.\label{psi-3D}\end{eqnarray}

The current density along the $z$-direction in the 3D case has in
the chosen gauge the usual form:

\begin{eqnarray}
\left\langle j_{z}\right\rangle  & = & \left.-\frac{ie_{0}\hbar}{2m_{c}}(\partial_{z}-\partial_{z'})\left\langle \psi\left(x,y,z,t\right)\psi^{*}\left(x,y,z',t\right)\right\rangle \right|_{z=z'}\,,\label{Jz-3D}\end{eqnarray}
 where \begin{eqnarray}
\psi(x,y,z,t) & = & \int\frac{dk}{2\pi}\int\frac{dq}{2\pi}\sum_{n\geq0}\psi_{q}(n,k,t)\cdot\mathrm{e}^{-iky}\mathrm{e}^{-iqz}u_{n}\left(x-\frac{\hbar k}{2eB}\right)\,,\label{FourierC-3D}\end{eqnarray}
 and the correlation function is, according to Eq. (\ref{psi-3D}),\begin{eqnarray}
\left\langle \psi(x,y,z,t)\psi^{*}(x,y,z',t)\right\rangle  & = & \frac{mk_{B}T}{\pi\hbar^{2}\xi_{0c}}h\int_{0}^{\infty}du\,\left.\left(\int\frac{dq'}{2\pi}\sum_{n}\right)\right|_{h(2n+1)+q'^{2}\leq c}\label{psi-corr-3D}\\
 &  & \cdot\exp\left[-i\frac{q'}{\xi_{0c}}\left(z-z'\right)-\frac{4p'^{2}u^{3}}{3}\right]\cdot\exp\left\{ -u\left[\widetilde{\varepsilon}+\left(2n+1\right)h+q'^{2}-2q'p'u\right]\right\} \,,\nonumber \end{eqnarray}
 where\begin{equation}
p'=\frac{\pi e\xi_{0c}}{16k_{B}T}\, E=\frac{\xi_{0c}\sqrt{3}}{\xi_{0}}\frac{E}{E_{0}}\,.\label{p-3D}\end{equation}

After careful evaluation of the $q'$ integral and the LL sum, with
consideration of the cut-off condition (\ref{Cut-off-3D}), one obtains
eventually the Hartree renormalization equation with both the magnetic
and electric fields applied along the $z$-direction, as:\begin{eqnarray}
\widetilde{\varepsilon}_{(E=E_{z},B=B_{z})}^{(3\mathrm{D})} & =\ln\frac{T}{T_{0}}+ & \frac{g^{(3\mathrm{D})}T}{\pi}\int_{0}^{\infty}du\,\frac{2uh\,\mathrm{e}^{-uh}}{1-\exp\left(-2uh\right)}\mathrm{e}^{-u\widetilde{\varepsilon}-\frac{4p'^{2}u^{3}}{3}}\int_{0}^{c+h}dw\,\mathrm{e}^{-uw}\frac{\sinh\left(2p'u^{2}\sqrt{w}\sqrt{\frac{c-h}{c+h}}\right)}{2p'u^{2}}\,,\label{epst-eq-zz-3D}\end{eqnarray}
\begin{equation}
g^{(3\mathrm{D})}=\frac{2\mu_{0}\kappa_{\mathrm{GL}}^{2}e^{2}\xi_{0}^{2}k_{B}}{\pi\hbar^{2}\xi_{0c}}\,,\label{g'}\end{equation}
 One should note that $\kappa_{\mathrm{GL}}$ in Eq. (\ref{g'}) is
the Ginzburg-Landau parameter in the $xy$-plane (and therefore proportional
to the Cooper pair mass $m$ in the $xy$-plane, which in turn is
proportional to $1/\xi_{0}^{2}$), so that the $g^{(3\mathrm{D})}$-parameter
is in fact proportional with the product $1/\xi_{0}^{2}\xi_{0c}$,
symmetric with respect to the coherence lengths along the three principal
axes of the material.

The general expression of the fluctuation conductivity (the AL contribution)
in a 3D anisotropic superconductor, when both the electric and magnetic
fields are applied along the symmetry axis, will write in turn:\begin{eqnarray}
\Delta\sigma_{zz}^{\mathrm{AL};(3\mathrm{D})}(E,B) & = & \frac{e^{2}\xi_{0c}}{8\pi\hbar\xi_{0}^{2}}\left(\frac{c-h}{c+h}\right)^{3/2}\int_{0}^{\infty}du\, u^{2}\frac{2uh\,\mathrm{e}^{-uh}}{1-\exp\left(-2uh\right)}\exp\left[-u\widetilde{\varepsilon}-\frac{4p'^{2}u^{3}}{3}\right]\label{sigmaZZ-3D}\\
 &  & \cdot\int_{0}^{c+h}dw\, w^{3/2}\mathrm{e}^{-uw}\left[\frac{\cosh\left(2p'u^{2}\sqrt{w}\sqrt{\frac{c-h}{c+h}}\right)}{\left(2p'u^{2}\sqrt{w}\sqrt{\frac{c-h}{c+h}}\right)^{2}}-\frac{\sinh\left(2p'u^{2}\sqrt{w}\sqrt{\frac{c-h}{c+h}}\right)}{\left(2p'u^{2}\sqrt{w}\sqrt{\frac{c-h}{c+h}}\right)^{3}}\right]\,.\nonumber \end{eqnarray}
 The above equation (\ref{sigmaZZ-3D}) takes a simpler form if one
neglects the cut-off (i.e. for $c\rightarrow\infty$), namely:\begin{eqnarray}
\left.\Delta\sigma_{zz}^{\mathrm{AL}}\right|_{\mathrm{NoCut}}^{(3\mathrm{D})}(E,B) & = & \frac{e^{2}\xi_{0c}}{32\sqrt{\pi}\hbar\xi_{0}^{2}}\int_{0}^{\infty}du\,\frac{2uh\,\mathrm{e}^{-uh}}{1-\exp\left(-2uh\right)}\frac{1}{\sqrt{u}}\exp\left(-u\widetilde{\varepsilon}-\frac{1}{3}p'^{2}u^{3}\right)\,.\label{sigmaZZ-3D-NoCut}\end{eqnarray}
\end{widetext}

This expression can be also directly inferred from the corresponding
result of the layered model, if one takes the 3D limit $s\rightarrow0,\: r\rightarrow\infty$
in Eq. (\ref{sigmazzC-noCut}). In this case $p\rightarrow0$, so
that $\sin(pu)/pu\rightarrow1$, while the sine function can be expanded
up to the cubic term in $u$ in the argument of the modified Bessel
function, which becomes thus $I_{1}\left(ru/2-\xi_{0c}^{2}E'^{2}u^{3}/\xi_{0}^{2}\right)$$\simeq$$\left(\pi r\right)^{-1/2}\exp\left(ru/2-p'^{2}u^{3}/3\right)$,
if one takes the asymptotic expression of the modified Bessel functions
$I_{n}(z)\simeq\mathrm{e}^{z}/\sqrt{2\pi z}$ for large arguments
$z\rightarrow\infty$. It must be reminded, however, that the general
expression (\ref{sigmaZZ-3D}), where the cut-off is considered, can
not be directly obtained from the corresponding result (\ref{sigmazz-C})
of the layered model, since for the latter the cut-off procedure is
not applied for the out-of-plane momentum.

\subsection{Limit cases of vanishing electric or/and magnetic field}

The linear response limit for the results (\ref{epst-eq-zz-3D}) and
(\ref{sigmaZZ-3D}), i.e. the case of a vanishing electric field and
a finite magnetic field, can be found directly from the correlation
function (\ref{psi-corr-3D}), by performing the $q'$ integral before
the LL sum. They write:\begin{widetext}

\begin{eqnarray}
\widetilde{\varepsilon}_{(E\rightarrow0,B>0)}^{(3\mathrm{D})} & = & \ln\frac{T}{T_{0}}+\frac{2g^{(3\mathrm{D})}T}{\pi}h\sum_{n=0}^{N_{c}}\frac{1}{\sqrt{\widetilde{\varepsilon}_{n}}}\arctan\sqrt{\frac{c+\widetilde{\varepsilon}}{\widetilde{\varepsilon}_{n}}-1}\,,\label{epst-eq-E0-3D}\\
\left.\Delta\sigma_{zz}^{\mathrm{AL}}\right|_{E\rightarrow0,B>0}^{(3\mathrm{D})} & = & \frac{e^{2}\xi_{0c}h}{16\pi\hbar\xi_{0}^{2}}\sum_{n=0}^{N_{c}}\left[\frac{1}{\widetilde{\varepsilon}_{n}^{3/2}}\arctan\sqrt{\frac{c+\widetilde{\varepsilon}}{\widetilde{\varepsilon}_{n}}-1}+\frac{\left(c+\widetilde{\varepsilon}-\widetilde{\varepsilon}_{n}\right)^{3/2}}{\widetilde{\varepsilon}_{n}\left(c+\widetilde{\varepsilon}\right)^{2}}-\frac{\left(c+\widetilde{\varepsilon}-\widetilde{\varepsilon}_{n}\right)^{1/2}}{\left(c+\widetilde{\varepsilon}\right)^{2}}\right]\,,\label{SigmaZZ-E0-3D}\end{eqnarray}
 where the LL cut-off index is $N_{c}=(c-h)/2h$ and $\widetilde{\varepsilon}_{n}=\widetilde{\varepsilon}+(2n+1)h$.

In the other limit of a vanishing magnetic field ($B=0$) but under
arbitrarily strong electric field ($E>0$), one obtains:

\begin{eqnarray}
\widetilde{\varepsilon}_{(E=E_{z},B=0)}^{(3\mathrm{D})} & = & \ln\frac{T}{T_{0}}+\frac{g^{(3\mathrm{D})}T}{\pi}\int_{0}^{\infty}du\,\exp\left(-u\widetilde{\varepsilon}-\frac{4p'^{2}u^{3}}{3}\right)\cdot\int_{0}^{c}dw\,\mathrm{e}^{-uw}\frac{\sinh\left(2p'u^{2}\sqrt{w}\right)}{2p'u^{2}}\,,\label{epst-eq-B0-zz-3D}\end{eqnarray}
 which is the 3D equivalent of the renormalization equation (\ref{selconsistCB0}),
and\begin{eqnarray}
\left.\Delta\sigma_{zz}^{\mathrm{AL}}\right|_{E>0,B=0}^{(3\mathrm{D})} & = & \frac{e^{2}\xi_{0c}}{8\pi\hbar\xi_{0}^{2}}\int_{0}^{\infty}du\, u^{2}\mathrm{e}^{-u\widetilde{\varepsilon}-\frac{4p'^{2}u^{3}}{3}}\int_{0}^{c}dw\, w^{3/2}\mathrm{e}^{-uw}\cdot\left[\frac{\cosh\left(2p'u^{2}\sqrt{w}\right)}{\left(2p'u^{2}\sqrt{w}\right)^{2}}-\frac{\sinh\left(2p'u^{2}\sqrt{w}\right)}{\left(2p'u^{2}\sqrt{w}\right)^{3}}\right]\,,\label{sigma-zz-3D}\end{eqnarray}
\end{widetext}

which is the 3D equivalent of the fluctuation conductivity (\ref{sigmazzB0})
from the layered model. Taken in the isotropic case ($\xi_{0c}=\xi_{0}$),
Eqs. (\ref{epst-eq-B0-zz-3D}) and (\ref{sigma-zz-3D}) match the
corresponding expressions already found in Ref. \onlinecite{PuicaLangE},
where the calculations were performed in 3D limit in connection to
the non-Ohmic \emph{in-plane} conductivity for the layered model.

If one neglected the cut-off ($c\rightarrow\infty$), the r.h.s term
in Eq. (\ref{epst-eq-B0-zz-3D}) would become divergent, while Eq.
(\ref{sigma-zz-3D}) would take the expression\begin{eqnarray}
\left.\Delta\sigma_{zz}^{\mathrm{AL}}\right|_{\mathrm{NoCut}}^{(3\mathrm{D})} & = & \frac{e^{2}\xi_{0c}}{32\sqrt{\pi}\hbar\xi_{0}^{2}}\int_{0}^{\infty}\frac{du}{\sqrt{u}}\mathrm{e}^{-\widetilde{\varepsilon}\, u-\frac{1}{3}p'^{2}u^{3}}\,,\label{SigmaZZ-3D-B0-NoCut}\end{eqnarray}
 already known \cite{Tsuzuki70,Dorsey91,Mishonov02} for isotropic
bulk supercondutors ($\xi_{0c}=\xi_{0}$) and Gaussian fluctuations
(i.e. with $\widetilde{\varepsilon}=\varepsilon$).

Taking further the limit $E\rightarrow0$, one obtains for the Hartree
renormalization equation:

\begin{equation}
\widetilde{\varepsilon}_{(E=0,B=0)}^{(3\mathrm{D})}=\ln\frac{T}{T_{0}}+\frac{2g^{(3\mathrm{D})}T}{\pi}\left(\sqrt{c}-\sqrt{\widetilde{\varepsilon}}\arctan\sqrt{\frac{c}{\widetilde{\varepsilon}}}\right)\,,\label{epst-E0B0-3D}\end{equation}
 and the fluctuation conductivity in the $z$ direction,

\begin{eqnarray}
\left.\Delta\sigma_{zz}^{\mathrm{AL}}\right|_{E=0,B=0}^{(3\mathrm{D})} & = & \frac{e^{2}\xi_{0c}}{48\pi\hbar\xi_{0}^{2}}\left[\frac{3\arctan\sqrt{\frac{c}{\widetilde{\varepsilon}}}}{\sqrt{\widetilde{\varepsilon}}}\right.\label{sigma-zz-E0B0-3D}\\
 &  & \left.-\frac{3\widetilde{\varepsilon}\sqrt{c}}{\left(c+\widetilde{\varepsilon}\right)^{2}}-\frac{5c^{3/2}}{\left(c+\widetilde{\varepsilon}\right)^{2}}\right]\,,\nonumber \end{eqnarray}
 The result (\ref{sigma-zz-E0B0-3D}) matches thus the expression
found \cite{Gauzzi95,Carballeira01} for Gaussian fluctuations ($\widetilde{\varepsilon}=\varepsilon$)
in a 3D isotropic superconductor ($\xi_{0c}=\xi_{0}$).

Equation (\ref{epst-E0B0-3D}) taken for $\widetilde{\varepsilon}=0$
gives the relation between the mean-field transition temperature $T_{0}$
and the actual critical temperature $T_{c0}=\left.T_{c}\right|_{E=0,B=0}$
where the superconductivity is attained in the absence of external
fields,\begin{equation}
T_{0}=T_{c0}\exp\left(2g^{(3\mathrm{D})}T_{c0}\sqrt{c}/\pi\right)\,,\label{T0-Tc0-3D}\end{equation}
 which represents the equivalent of Eq. (\ref{T0-Tc0}) from the layered
model.

\subsection{Estimation of DOS term}

Supposing that the same proportionality (\ref{proportDOS}) between
the DOS contribution to the fluctuation conductivity $\Delta\sigma_{zz}^{\mathrm{DOS}}$
and the Cooper pairs density $\left\langle \left|\psi\right|^{2}\right\rangle $
holds also in the 3D case, one can estimate for the different combinations
of electric and magnetic fields,\begin{widetext}\begin{eqnarray}
\left.\Delta\sigma_{zz}^{\mathrm{DOS}}(E,B)\right|^{(3\mathrm{D})} & = & -\frac{e^{2}\xi_{0c}\kappa}{4\pi\hbar\xi_{0}^{2}}\int_{0}^{\infty}du\,\frac{2uh\,\mathrm{e}^{-uh}}{1-\exp\left(-2uh\right)}\mathrm{e}^{-u\widetilde{\varepsilon}-\frac{4p'^{2}u^{3}}{3}}\int_{0}^{c+h}dw\,\mathrm{e}^{-uw}\frac{\sinh\left(2p'u^{2}\sqrt{w}\sqrt{\frac{c-h}{c+h}}\right)}{2p'u^{2}}\,,\label{sigmaZZ-DOS-3D-EB}\\
\left.\Delta\sigma_{zz}^{\mathrm{DOS}}(B)\right|_{E\rightarrow0}^{(3\mathrm{D})} & = & -\frac{e^{2}\xi_{0c}\kappa\, h}{2\pi\hbar\xi_{0}^{2}}\sum_{n=0}^{N_{c}}\frac{1}{\sqrt{\widetilde{\varepsilon}_{n}}}\arctan\sqrt{\frac{c+\widetilde{\varepsilon}}{\widetilde{\varepsilon}_{n}}-1}\,,\label{sigmaZZ-DOS-3D}\\
\left.\Delta\sigma_{zz}^{\mathrm{DOS}}(E)\right|_{B=0}^{(3\mathrm{D})} & = & -\frac{e^{2}\xi_{0c}\kappa}{4\pi\hbar\xi_{0}^{2}}\int_{0}^{\infty}du\,\exp\left(-u\widetilde{\varepsilon}-\frac{4p'^{2}u^{3}}{3}\right)\cdot\int_{0}^{c}dw\,\mathrm{e}^{-uw}\frac{\sinh\left(2p'u^{2}\sqrt{w}\right)}{2p'u^{2}}\,,\label{sigmaZZ-DOS-3D-E}\\
\left.\Delta\sigma_{zz}^{\mathrm{DOS}}\right|_{B=0,E\rightarrow0}^{(3\mathrm{D})} & = & -\frac{e^{2}\xi_{0c}\kappa}{2\pi\hbar\xi_{0}^{2}}\left(\sqrt{c}-\sqrt{\widetilde{\varepsilon}}\arctan\sqrt{\frac{c}{\widetilde{\varepsilon}}}\right)\,.\label{sigmaZZ-DOS-3D-E0B0}\end{eqnarray}
\end{widetext}

Unlike the layered model, where, for instance, the AL contribution
(\ref{sigmazzE0}) was quadratic in the anisotropy parameter $r$
while the DOS one (\ref{DorinDOS}) was linear, the corresponding
3D fluctuation contributions (\ref{SigmaZZ-E0-3D}) and (\ref{sigmaZZ-DOS-3D})
are both proportional to the ratio $\xi_{0c}/\xi_{0}^{2}$, while
the AL one is more singular in $\widetilde{\varepsilon}$. The DOS
contribution might thus be of less importance in the 3D case with
respect to the layered model. However, the relations (\ref{sigmaZZ-DOS-3D-EB}),
(\ref{sigmaZZ-DOS-3D}), (\ref{sigmaZZ-DOS-3D-E}) and (\ref{sigmaZZ-DOS-3D-E0B0})
remain to be confirmed or refuted by a microscopic approach.

\subsection{Arbitrary orientation of the electric field}

The 3D equivalents of Eqs. (\ref{epstilde-Etilted}), (\ref{jx-Etilted}),
(\ref{jz-Etilted}), (\ref{jx-Etilted-NoCut}) and (\ref{jz-Etilted-NoCut}),
valid for an arbitrary orientation of the electric field with respect
to the layers, can be obtained in a simpler manner, without having
to solve again the TDGL equation, by using a special scaling transformation
of the coordinates and field components that reduces the problem to
the isotropic case.\cite{Blatter92,Larkin05} In the isotropic system,
the coordinate axes can be in turn freely rotated so that the electric
field acquires again only one non-zero component, for which the solution
is already known. For the anisotropic 3D model with axial symmetry
treated in this Section, having the zero-temperature-extrapolated
coherence lengths $\xi_{0c}$ in the $z$-direction and $\xi_{0}$
in the $xy$-plane, corresponding to the anisotropy factor\begin{equation}
\gamma_{a}=\frac{\xi_{0c}}{\xi_{0}}=\sqrt{\frac{m}{m_{c}}}<1\,.\label{gammaA}\end{equation}
 the scaling transformation of the coordinates and vector potential
\cite{Blatter92,Larkin05} writes:\begin{eqnarray}
\widetilde{x}=x; & \; & \widetilde{y}=y;\quad\widetilde{z}=\frac{z}{\gamma_{a}};\label{scaling}\\
\widetilde{A_{x}}=A_{x}; & \; & \widetilde{A_{y}}=A_{y};\quad\widetilde{A_{z}}=\gamma_{a}A_{z}\,,\nonumber \end{eqnarray}
 This transformation implies the rescaling of the fields as\begin{eqnarray}
\widetilde{B_{x}}=\gamma_{a}B_{x}; & \; & \widetilde{B_{y}}=\gamma_{a}B_{y};\quad\widetilde{B_{z}}=B_{z};\nonumber \\
\widetilde{E_{x}}=E_{x}; & \; & \widetilde{E_{y}}=E_{y};\quad\widetilde{E_{z}}=\gamma_{a}E_{z};\label{FieldScaling}\end{eqnarray}
 and for the order parameter and current density as\begin{eqnarray}
\left|\widetilde{\psi}\right|^{2} & = & \gamma_{a}\left|\psi\right|^{2};\label{CurrentScaling}\\
\widetilde{j_{x}}=\gamma_{a}j_{x}; & \; & \widetilde{j_{y}}=\gamma_{a}j_{y};\quad\widetilde{j_{z}}=j_{z}\,.\nonumber \end{eqnarray}
 Thus, the anisotropy is removed from the gradient terms in the GL
equation and reintroduced into the magnetic energy term, whose fluctuation
can however be usually neglected for hard type-II superconductors.\cite{Blatter92}

In the case of a zero magnetic field and an electric field $\left(E_{x}=E\sin\theta,\, E_{y}=0,\, E_{z}=E\cos\theta\right)$
applied in the anisotropic material at an angle $\theta$ with the
$z$-axis, the problem can be reduced, according to Eqs. (\ref{FieldScaling}),
to that of an electric field $\left(\widetilde{E_{x}}=\widetilde{E}\sin\widetilde{\theta},\,\widetilde{E_{y}}=0,\,\widetilde{E_{z}}=\widetilde{E}\cos\widetilde{\theta}\right)$
applied on an isotropic superconductor having the coherence length
$\xi_{0}$, such as:\begin{eqnarray}
\widetilde{E} & = & E\left(\sin^{2}\theta+\gamma_{a}^{2}\cos^{2}\theta\right)^{1/2}\,,\nonumber \\
\tan\widetilde{\theta} & = & \frac{1}{\gamma_{a}}\tan\theta\,.\label{Etilde-Etilted}\end{eqnarray}

Taking now into account Eqs. (\ref{epst-eq-B0-zz-3D}), (\ref{sigma-zz-3D})
and (\ref{SigmaZZ-3D-B0-NoCut}) considered for $\xi_{0c}=\xi_{0}$,
together with Eqs. (\ref{CurrentScaling}), one obtains eventually\begin{widetext}\begin{eqnarray}
\widetilde{\varepsilon}_{(E_{x},E_{z})}^{(3\mathrm{D})} & = & \ln\frac{T}{T_{0}}+\frac{g^{(3\mathrm{D})}T}{\pi}\int_{0}^{\infty}du\,\exp\left[-u\widetilde{\varepsilon}-4\left(\frac{\widetilde{E}}{E_{0}}\right)^{2}u^{3}\right]\cdot\int_{0}^{c}dw\,\mathrm{e}^{-uw}\frac{\sinh\left(2\sqrt{3}\frac{\widetilde{E}}{E_{0}}u^{2}\sqrt{w}\right)}{2\sqrt{3}\frac{\widetilde{E}}{E_{0}}u^{2}}\,,\label{Epstiled-Etilted-3D}\\
\Delta j_{x}^{(3\mathrm{D})} & = & E_{x}\frac{e^{2}}{8\pi\hbar\xi_{0c}}\int_{0}^{\infty}du\, u^{2}\mathrm{e}^{-u\widetilde{\varepsilon}-4\left(\frac{\widetilde{E}}{E_{0}}\right)^{2}u^{3}}\int_{0}^{c}dw\, w^{3/2}\mathrm{e}^{-uw}\cdot\left[\frac{\cosh\left(2\sqrt{3}\frac{\widetilde{E}}{E_{0}}u^{2}\sqrt{w}\right)}{\left(2\sqrt{3}\frac{\widetilde{E}}{E_{0}}u^{2}\sqrt{w}\right)^{2}}-\frac{\sinh\left(2\sqrt{3}\frac{\widetilde{E}}{E_{0}}u^{2}\sqrt{w}\right)}{\left(2\sqrt{3}\frac{\widetilde{E}}{E_{0}}u^{2}\sqrt{w}\right)^{3}}\right]\,,\label{jx-Etilted-3D}\\
\Delta j_{z}^{(3\mathrm{D})} & = & E_{z}\frac{e^{2}\xi_{0c}}{8\pi\hbar\xi_{0}^{2}}\int_{0}^{\infty}du\, u^{2}\mathrm{e}^{-u\widetilde{\varepsilon}-4\left(\frac{\widetilde{E}}{E_{0}}\right)^{2}u^{3}}\int_{0}^{c}dw\, w^{3/2}\mathrm{e}^{-uw}\cdot\left[\frac{\cosh\left(2\sqrt{3}\frac{\widetilde{E}}{E_{0}}u^{2}\sqrt{w}\right)}{\left(2\sqrt{3}\frac{\widetilde{E}}{E_{0}}u^{2}\sqrt{w}\right)^{2}}-\frac{\sinh\left(2\sqrt{3}\frac{\widetilde{E}}{E_{0}}u^{2}\sqrt{w}\right)}{\left(2\sqrt{3}\frac{\widetilde{E}}{E_{0}}u^{2}\sqrt{w}\right)^{3}}\right]\,,\label{jz-Etilted-3D}\end{eqnarray}
 so that neglecting the cut-off,\begin{eqnarray}
\left.\Delta j_{x}^{(3\mathrm{D})}\right|_{\mathrm{NoCut}} & = & E_{x}\cdot\frac{e^{2}}{32\sqrt{\pi}\hbar\xi_{0c}}\int_{0}^{\infty}\frac{du}{\sqrt{u}}\exp\left[-\widetilde{\varepsilon}\, u-\left(\frac{\widetilde{E}}{E_{0}}\right)^{2}u^{3}\right]\,,\label{jx-Etilted-3D-NoCut}\\
\left.\Delta j_{z}^{(3\mathrm{D})}\right|_{\mathrm{NoCut}} & = & E_{z}\cdot\frac{e^{2}\xi_{0c}}{32\sqrt{\pi}\hbar\xi_{0}^{2}}\int_{0}^{\infty}\frac{du}{\sqrt{u}}\exp\left[-\widetilde{\varepsilon}\, u-\left(\frac{\widetilde{E}}{E_{0}}\right)^{2}u^{3}\right]\,.\label{jz-Etilted-3D-NoCut}\end{eqnarray}
\end{widetext}

\section{Conclusion}

In this work we have theoretically investigated the non-Ohmic effect
of an arbitrarily strong electric field on the out-of-plane fluctuation
magneto-conductivity of a layered superconductor. Our framework was
provided by the Langevin approach to the TDGL equation, and the Hartree
approximation was used in order to take into account the fluctuation
interaction. The main general results of our work, valid when magnetic
and electric fields of arbitrary magnitude are applied perpendicular
to the layers, are the formulae (\ref{self-consist-C}) for the renormalized
reduced temperature and, respectively, (\ref{sigmazz-C}) for the
AL contribution to the out-of-plane fluctuation conductivity, as well
as the estimation (\ref{sigmazzDOS}) for the DOS term. In the limit
case of a vanishing electric field, the results were found to reduce
to the expressions already known from the linear response approximation
(Section \ref{LimitCases}). Extensions of the results have been provided
for the case of a tilted electric field with respect to the crystallographic
axes (Section \ref{TiltedE}), as well as for 3D anisotropic superconductors
(Section \ref{Non-layered}).

In order to illustrate the predictions of the theoretical calculations,
we have taken as example a typical HTSC material, like the optimally
doped YBa$_{2}$Cu$_{3}$O$_{6+x}$, and evidenced the non-linear
effect of a strong electric field through comparison to the results
obtained in the linear response approximation, in the presence of
a finite or of a zero magnetic field. An important fluctuation suppression
in the AL part of the out-of-plane conductivity has been predicted
for electric fields of hundreds of V/cm, while the non-Ohmic effect
on the DOS contribution turned out to be marginal in the same range
(Fig. \ref{FigRhozz}).

So far, the effect of a strong electric field on the transport properties
of HTSC has been investigated and proven experimentally only for the
in-plane fluctuation conductivity and only in the absence of magnetic
field.\cite{Soret93,Gorlova95,Kunchur95,pssc05} The difficulties
lie in the high dissipation in cuprates (of the order of $\mathrm{GWcm}^{-3}$
at electric fields of hundreds V/cm) that can increase the sample
temperature to values where the non-linearity is no longer discernable.
In this connection, applying short current pulses (tens of ns) at
high current densities (a few $\mathrm{MAcm}^{-2}$), in combination
with using very thin films (under 50 nm thick) in order to enhance
the heat transport to the substrate,\cite{pssc05} seems to be a better
alternative to the dc and ac measurements. This procedure, if applied
on vicinal films in order to access also the $c$-axis conduction,
will probably allow for the necessary accuracy necessary to detect
the non-Ohmic behavior also for the out-of-plane fluctuation conductivity.

\begin{acknowledgments}
This work was supported by the Austrian Fonds zur F\"{o}rderung der
wissenschaftlichen Forschung and the Micro@Nanofabrication (MNA) Network,
funded by the Austrian Ministry for Economic Affairs and Labour.
\end{acknowledgments}
\bibliographystyle{APSREV}
\bibliography{PuicaLang4}

\begin{thebibliography}{39}
\expandafter\ifx\csname natexlab\endcsname\relax\def\natexlab#1{#1}\fi
\expandafter\ifx\csname bibnamefont\endcsname\relax
  \def\bibnamefont#1{#1}\fi
\expandafter\ifx\csname bibfnamefont\endcsname\relax
  \def\bibfnamefont#1{#1}\fi
\expandafter\ifx\csname citenamefont\endcsname\relax
  \def\citenamefont#1{#1}\fi
\expandafter\ifx\csname url\endcsname\relax
  \def\url#1{\texttt{#1}}\fi
\expandafter\ifx\csname urlprefix\endcsname\relax\def\urlprefix{URL }\fi
\providecommand{\bibinfo}[2]{#2}
\providecommand{\eprint}[2][]{\url{#2}}

\bibitem[{\citenamefont{Aslamazov and Larkin}(1968)}]{Aslamazov68}
\bibinfo{author}{\bibfnamefont{L.}~\bibnamefont{Aslamazov}} \bibnamefont{and}
  \bibinfo{author}{\bibfnamefont{A.}~\bibnamefont{Larkin}},
  \bibinfo{journal}{Phys. Lett.} \textbf{\bibinfo{volume}{26A}},
  \bibinfo{pages}{238} (\bibinfo{year}{1968}).

\bibitem[{\citenamefont{Lawrence and Doniach}(1971)}]{Lawrence71}
\bibinfo{author}{\bibfnamefont{W.~E.} \bibnamefont{Lawrence}} \bibnamefont{and}
  \bibinfo{author}{\bibfnamefont{S.}~\bibnamefont{Doniach}}, in
  \emph{\bibinfo{booktitle}{Proc. of the 12th Int. Conf. on Low-Temp. Phys.,
  Kyoto, Japan, 1970}}, edited by
  \bibinfo{editor}{\bibfnamefont{E.}~\bibnamefont{Kanda}}
  (\bibinfo{publisher}{Keigaku, Tokyo}, \bibinfo{year}{1971}), p.
  \bibinfo{pages}{361}.

\bibitem[{\citenamefont{Maki}(1969)}]{Maki69}
\bibinfo{author}{\bibfnamefont{K.}~\bibnamefont{Maki}}, \bibinfo{journal}{J.
  Low. Temp. Phys.} \textbf{\bibinfo{volume}{1}}, \bibinfo{pages}{513}
  (\bibinfo{year}{1969}).

\bibitem[{\citenamefont{Klemm}(1974)}]{Klemm74}
\bibinfo{author}{\bibfnamefont{R.~A.} \bibnamefont{Klemm}},
  \bibinfo{journal}{J. Low. Temp. Phys.} \textbf{\bibinfo{volume}{16}},
  \bibinfo{pages}{381} (\bibinfo{year}{1974}).

\bibitem[{\citenamefont{Fukuyama et~al.}(1971)\citenamefont{Fukuyama, Ebisawa,
  and Tsuzuki}}]{FET}
\bibinfo{author}{\bibfnamefont{H.}~\bibnamefont{Fukuyama}},
  \bibinfo{author}{\bibfnamefont{H.}~\bibnamefont{Ebisawa}}, \bibnamefont{and}
  \bibinfo{author}{\bibfnamefont{T.}~\bibnamefont{Tsuzuki}},
  \bibinfo{journal}{Prog. Theor. Phys.} \textbf{\bibinfo{volume}{46}},
  \bibinfo{pages}{1028} (\bibinfo{year}{1971}).

\bibitem[{\citenamefont{Dorin et~al.}(1993)\citenamefont{Dorin, Klemm,
  Varlamov, Buzdin, and Livanov}}]{Dorin93}
\bibinfo{author}{\bibfnamefont{V.~V.} \bibnamefont{Dorin}},
  \bibinfo{author}{\bibfnamefont{R.~A.} \bibnamefont{Klemm}},
  \bibinfo{author}{\bibfnamefont{A.~A.} \bibnamefont{Varlamov}},
  \bibinfo{author}{\bibfnamefont{A.~I.} \bibnamefont{Buzdin}},
  \bibnamefont{and} \bibinfo{author}{\bibfnamefont{D.~V.}
  \bibnamefont{Livanov}}, \bibinfo{journal}{Phys. Rev. B}
  \textbf{\bibinfo{volume}{48}}, \bibinfo{pages}{12951} (\bibinfo{year}{1993}).

\bibitem[{\citenamefont{Ikeda et~al.}(1991)\citenamefont{Ikeda, Ohmi, and
  Tsuneto}}]{Ikeda91a}
\bibinfo{author}{\bibfnamefont{R.}~\bibnamefont{Ikeda}},
  \bibinfo{author}{\bibfnamefont{T.}~\bibnamefont{Ohmi}}, \bibnamefont{and}
  \bibinfo{author}{\bibfnamefont{T.}~\bibnamefont{Tsuneto}},
  \bibinfo{journal}{J. Phys. Soc. Jap.} \textbf{\bibinfo{volume}{60}},
  \bibinfo{pages}{1051} (\bibinfo{year}{1991}), \bibinfo{note}{and references
  therein}.

\bibitem[{\citenamefont{Ullah and Dorsey}(1991)}]{UD}
\bibinfo{author}{\bibfnamefont{S.}~\bibnamefont{Ullah}} \bibnamefont{and}
  \bibinfo{author}{\bibfnamefont{A.~T.} \bibnamefont{Dorsey}},
  \bibinfo{journal}{Phys. Rev. B} \textbf{\bibinfo{volume}{44}},
  \bibinfo{pages}{262} (\bibinfo{year}{1991}).

\bibitem[{\citenamefont{Nishio and Ebisawa}(1997)}]{NE}
\bibinfo{author}{\bibfnamefont{T.}~\bibnamefont{Nishio}} \bibnamefont{and}
  \bibinfo{author}{\bibfnamefont{H.}~\bibnamefont{Ebisawa}},
  \bibinfo{journal}{Physica C} \textbf{\bibinfo{volume}{290}},
  \bibinfo{pages}{43} (\bibinfo{year}{1997}).

\bibitem[{\citenamefont{Livanov et~al.}(1997)\citenamefont{Livanov, Milani,
  Balestrino, and Aruta}}]{Livanov97}
\bibinfo{author}{\bibfnamefont{D.~V.} \bibnamefont{Livanov}},
  \bibinfo{author}{\bibfnamefont{E.}~\bibnamefont{Milani}},
  \bibinfo{author}{\bibfnamefont{G.}~\bibnamefont{Balestrino}},
  \bibnamefont{and} \bibinfo{author}{\bibfnamefont{C.}~\bibnamefont{Aruta}},
  \bibinfo{journal}{Phys. Rev. B} \textbf{\bibinfo{volume}{55}},
  \bibinfo{pages}{R8701} (\bibinfo{year}{1997}).

\bibitem[{\citenamefont{Hurault}(1969)}]{Hurault69}
\bibinfo{author}{\bibfnamefont{J.~P.} \bibnamefont{Hurault}},
  \bibinfo{journal}{Phys. Rev.} \textbf{\bibinfo{volume}{179}},
  \bibinfo{pages}{494} (\bibinfo{year}{1969}).

\bibitem[{\citenamefont{Larkin and Varlamov}(2005)}]{Larkin05}
\bibinfo{author}{\bibfnamefont{A.~I.} \bibnamefont{Larkin}} \bibnamefont{and}
  \bibinfo{author}{\bibfnamefont{A.}~\bibnamefont{Varlamov}},
  \emph{\bibinfo{title}{Theory of Fluctuations in Superconductors}}, no.
  \bibinfo{number}{127} in \bibinfo{series}{International Series of Monographs
  on Physics} (\bibinfo{publisher}{Clarendon Press, Oxford},
  \bibinfo{year}{2005}).

\bibitem[{\citenamefont{Schmid}(1969)}]{Schmid69}
\bibinfo{author}{\bibfnamefont{A.}~\bibnamefont{Schmid}},
  \bibinfo{journal}{Phys. Rev.} \textbf{\bibinfo{volume}{180}},
  \bibinfo{pages}{527} (\bibinfo{year}{1969}).

\bibitem[{\citenamefont{Tsuzuki}(1970)}]{Tsuzuki70}
\bibinfo{author}{\bibfnamefont{T.}~\bibnamefont{Tsuzuki}},
  \bibinfo{journal}{Progr. Theoret. Phys. (Kyoto)}
  \textbf{\bibinfo{volume}{43}}, \bibinfo{pages}{286} (\bibinfo{year}{1970}).

\bibitem[{\citenamefont{Thomas and Parks}(1971)}]{Thomas71}
\bibinfo{author}{\bibfnamefont{G.~A.} \bibnamefont{Thomas}} \bibnamefont{and}
  \bibinfo{author}{\bibfnamefont{R.~D.} \bibnamefont{Parks}},
  \bibinfo{journal}{Physica} \textbf{\bibinfo{volume}{55}},
  \bibinfo{pages}{215} (\bibinfo{year}{1971}).

\bibitem[{\citenamefont{Kajimura et~al.}(1971)\citenamefont{Kajimura,
  Mikoshiba, and Yamaji}}]{Kajimura71}
\bibinfo{author}{\bibfnamefont{K.}~\bibnamefont{Kajimura}},
  \bibinfo{author}{\bibfnamefont{N.}~\bibnamefont{Mikoshiba}},
  \bibnamefont{and} \bibinfo{author}{\bibfnamefont{K.}~\bibnamefont{Yamaji}},
  \bibinfo{journal}{Phys. Rev. B} \textbf{\bibinfo{volume}{4}},
  \bibinfo{pages}{209} (\bibinfo{year}{1971}).

\bibitem[{\citenamefont{Varlamov and Reggiani}(1992)}]{Varlamov92}
\bibinfo{author}{\bibfnamefont{A.~A.} \bibnamefont{Varlamov}} \bibnamefont{and}
  \bibinfo{author}{\bibfnamefont{L.}~\bibnamefont{Reggiani}},
  \bibinfo{journal}{Phys. Rev. B} \textbf{\bibinfo{volume}{45}},
  \bibinfo{pages}{R1060} (\bibinfo{year}{1992}).

\bibitem[{\citenamefont{Mishonov et~al.}(2002)\citenamefont{Mishonov,
  Posazhennikova, and Indekeu}}]{Mishonov02}
\bibinfo{author}{\bibfnamefont{T.}~\bibnamefont{Mishonov}},
  \bibinfo{author}{\bibfnamefont{A.}~\bibnamefont{Posazhennikova}},
  \bibnamefont{and} \bibinfo{author}{\bibfnamefont{J.}~\bibnamefont{Indekeu}},
  \bibinfo{journal}{Phys. Rev. B} \textbf{\bibinfo{volume}{65}},
  \bibinfo{pages}{064519} (\bibinfo{year}{2002}).

\bibitem[{\citenamefont{Puica and Lang}(2003{\natexlab{a}})}]{PuicaLangE}
\bibinfo{author}{\bibfnamefont{I.}~\bibnamefont{Puica}} \bibnamefont{and}
  \bibinfo{author}{\bibfnamefont{W.}~\bibnamefont{Lang}},
  \bibinfo{journal}{Phys. Rev. B} \textbf{\bibinfo{volume}{68}},
  \bibinfo{pages}{054517} (\bibinfo{year}{2003}{\natexlab{a}}).

\bibitem[{\citenamefont{Soret et~al.}(1993)\citenamefont{Soret, Ammor,
  Martinie, Lecomte, Odier, and Bok}}]{Soret93}
\bibinfo{author}{\bibfnamefont{J.~C.} \bibnamefont{Soret}},
  \bibinfo{author}{\bibfnamefont{L.}~\bibnamefont{Ammor}},
  \bibinfo{author}{\bibfnamefont{B.}~\bibnamefont{Martinie}},
  \bibinfo{author}{\bibfnamefont{J.}~\bibnamefont{Lecomte}},
  \bibinfo{author}{\bibfnamefont{P.}~\bibnamefont{Odier}}, \bibnamefont{and}
  \bibinfo{author}{\bibfnamefont{J.}~\bibnamefont{Bok}},
  \bibinfo{journal}{Europhys. Lett.} \textbf{\bibinfo{volume}{21}},
  \bibinfo{pages}{617} (\bibinfo{year}{1993}).

\bibitem[{\citenamefont{Gorlova et~al.}(1995)\citenamefont{Gorlova, Zybtsev,
  and Pokrovskii}}]{Gorlova95}
\bibinfo{author}{\bibfnamefont{I.~G.} \bibnamefont{Gorlova}},
  \bibinfo{author}{\bibfnamefont{S.~G.} \bibnamefont{Zybtsev}},
  \bibnamefont{and} \bibinfo{author}{\bibfnamefont{V.~I.}
  \bibnamefont{Pokrovskii}}, \bibinfo{journal}{JETP Lett.}
  \textbf{\bibinfo{volume}{61}}, \bibinfo{pages}{839} (\bibinfo{year}{1995}).

\bibitem[{\citenamefont{Kunchur}(1995)}]{Kunchur95}
\bibinfo{author}{\bibfnamefont{M.~N.} \bibnamefont{Kunchur}},
  \bibinfo{journal}{Mod. Phys. Lett. B} \textbf{\bibinfo{volume}{9}},
  \bibinfo{pages}{399} (\bibinfo{year}{1995}).

\bibitem[{\citenamefont{Fruchter et~al.}(2004)\citenamefont{Fruchter, Sfar,
  Bouquet, Li, and Raffy}}]{Fruchter04}
\bibinfo{author}{\bibfnamefont{L.}~\bibnamefont{Fruchter}},
  \bibinfo{author}{\bibfnamefont{I.}~\bibnamefont{Sfar}},
  \bibinfo{author}{\bibfnamefont{F.}~\bibnamefont{Bouquet}},
  \bibinfo{author}{\bibfnamefont{Z.~Z.} \bibnamefont{Li}}, \bibnamefont{and}
  \bibinfo{author}{\bibfnamefont{H.}~\bibnamefont{Raffy}},
  \bibinfo{journal}{Phys. Rev. B} \textbf{\bibinfo{volume}{69}},
  \bibinfo{pages}{144511} (\bibinfo{year}{2004}).

\bibitem[{\citenamefont{Lang et~al.}(2005)\citenamefont{Lang, Puica, Peruzzi,
  Lemmermann, Pedarnig, and Bäuerle}}]{pssc05}
\bibinfo{author}{\bibfnamefont{W.}~\bibnamefont{Lang}},
  \bibinfo{author}{\bibfnamefont{I.}~\bibnamefont{Puica}},
  \bibinfo{author}{\bibfnamefont{M.}~\bibnamefont{Peruzzi}},
  \bibinfo{author}{\bibfnamefont{L.}~\bibnamefont{Lemmermann}},
  \bibinfo{author}{\bibfnamefont{J.~D.} \bibnamefont{Pedarnig}},
  \bibnamefont{and} \bibinfo{author}{\bibfnamefont{D.}~\bibnamefont{Bäuerle}},
  \bibinfo{journal}{phys. stat. sol. (c)} \textbf{\bibinfo{volume}{2}},
  \bibinfo{pages}{1615} (\bibinfo{year}{2005}).

\bibitem[{\citenamefont{Puica and Lang}(2003{\natexlab{b}})}]{PuicaLangM}
\bibinfo{author}{\bibfnamefont{I.}~\bibnamefont{Puica}} \bibnamefont{and}
  \bibinfo{author}{\bibfnamefont{W.}~\bibnamefont{Lang}},
  \bibinfo{journal}{Phys. Rev. B} \textbf{\bibinfo{volume}{68}},
  \bibinfo{pages}{212503} (\bibinfo{year}{2003}{\natexlab{b}}).

\bibitem[{\citenamefont{Puica and Lang}(2004)}]{PuicaLangH}
\bibinfo{author}{\bibfnamefont{I.}~\bibnamefont{Puica}} \bibnamefont{and}
  \bibinfo{author}{\bibfnamefont{W.}~\bibnamefont{Lang}},
  \bibinfo{journal}{Phys. Rev. B} \textbf{\bibinfo{volume}{70}},
  \bibinfo{pages}{092507} (\bibinfo{year}{2004}).

\bibitem[{\citenamefont{Pedarnig et~al.}(2002)\citenamefont{Pedarnig, Rössler,
  Delamare, Lang, Bäuerle, Köhler, and Zandbergen}}]{Pedarnig02}
\bibinfo{author}{\bibfnamefont{J.~D.} \bibnamefont{Pedarnig}},
  \bibinfo{author}{\bibfnamefont{R.}~\bibnamefont{Rössler}},
  \bibinfo{author}{\bibfnamefont{M.~P.} \bibnamefont{Delamare}},
  \bibinfo{author}{\bibfnamefont{W.}~\bibnamefont{Lang}},
  \bibinfo{author}{\bibfnamefont{D.}~\bibnamefont{Bäuerle}},
  \bibinfo{author}{\bibfnamefont{A.}~\bibnamefont{Köhler}}, \bibnamefont{and}
  \bibinfo{author}{\bibfnamefont{H.~W.} \bibnamefont{Zandbergen}},
  \bibinfo{journal}{Appl. Phys. Lett.} \textbf{\bibinfo{volume}{81}},
  \bibinfo{pages}{2587} (\bibinfo{year}{2002}).

\bibitem[{\citenamefont{Masker et~al.}(1969)\citenamefont{Masker, Marcelja, and
  Parks}}]{Masker69}
\bibinfo{author}{\bibfnamefont{W.~E.} \bibnamefont{Masker}},
  \bibinfo{author}{\bibfnamefont{S.}~\bibnamefont{Marcelja}}, \bibnamefont{and}
  \bibinfo{author}{\bibfnamefont{R.~D.} \bibnamefont{Parks}},
  \bibinfo{journal}{Phys. Rev.} \textbf{\bibinfo{volume}{188}},
  \bibinfo{pages}{745} (\bibinfo{year}{1969}).

\bibitem[{\citenamefont{Cyrot}(1973)}]{Cyrot73}
\bibinfo{author}{\bibfnamefont{M.}~\bibnamefont{Cyrot}}, \bibinfo{journal}{Rep.
  Prog. Phys.} \textbf{\bibinfo{volume}{36}}, \bibinfo{pages}{103}
  (\bibinfo{year}{1973}).

\bibitem[{\citenamefont{Mishonov and Penev}(2000)}]{Penev0}
\bibinfo{author}{\bibfnamefont{T.}~\bibnamefont{Mishonov}} \bibnamefont{and}
  \bibinfo{author}{\bibfnamefont{E.}~\bibnamefont{Penev}},
  \bibinfo{journal}{Int. J. Mod. Phys. B} \textbf{\bibinfo{volume}{14}},
  \bibinfo{pages}{3831} (\bibinfo{year}{2000}).

\bibitem[{\citenamefont{Ramallo}(2004)}]{Ramallo04}
\bibinfo{author}{\bibfnamefont{M.~V.} \bibnamefont{Ramallo}},
  \bibinfo{journal}{Europhys. Lett.} \textbf{\bibinfo{volume}{65}},
  \bibinfo{pages}{249} (\bibinfo{year}{2004}).

\bibitem[{\citenamefont{Vidal et~al.}(2002)\citenamefont{Vidal, Carballeira,
  Curràs, Mosqueira, Ramallo, Veira, and Viña}}]{Vidal02}
\bibinfo{author}{\bibfnamefont{F.}~\bibnamefont{Vidal}},
  \bibinfo{author}{\bibfnamefont{C.}~\bibnamefont{Carballeira}},
  \bibinfo{author}{\bibfnamefont{S.~R.} \bibnamefont{Curràs}},
  \bibinfo{author}{\bibfnamefont{J.}~\bibnamefont{Mosqueira}},
  \bibinfo{author}{\bibfnamefont{M.~V.} \bibnamefont{Ramallo}},
  \bibinfo{author}{\bibfnamefont{J.~A.} \bibnamefont{Veira}}, \bibnamefont{and}
  \bibinfo{author}{\bibfnamefont{J.}~\bibnamefont{Viña}},
  \bibinfo{journal}{Europhys. Lett.} \textbf{\bibinfo{volume}{59}},
  \bibinfo{pages}{754} (\bibinfo{year}{2002}).

\bibitem[{\citenamefont{Soto et~al.}(2004)\citenamefont{Soto, Carballeira,
  Mosqueira, Ramallo, Ruibal, Veira, and Vidal}}]{Soto04}
\bibinfo{author}{\bibfnamefont{F.}~\bibnamefont{Soto}},
  \bibinfo{author}{\bibfnamefont{C.}~\bibnamefont{Carballeira}},
  \bibinfo{author}{\bibfnamefont{J.}~\bibnamefont{Mosqueira}},
  \bibinfo{author}{\bibfnamefont{M.~V.} \bibnamefont{Ramallo}},
  \bibinfo{author}{\bibfnamefont{M.}~\bibnamefont{Ruibal}},
  \bibinfo{author}{\bibfnamefont{J.~A.} \bibnamefont{Veira}}, \bibnamefont{and}
  \bibinfo{author}{\bibfnamefont{F.}~\bibnamefont{Vidal}},
  \bibinfo{journal}{Phys. Rev. B} \textbf{\bibinfo{volume}{70}},
  \bibinfo{pages}{060501(R)} (\bibinfo{year}{2004}).

\bibitem[{\citenamefont{Levin}(2004)}]{Levin04}
\bibinfo{author}{\bibfnamefont{G.~A.} \bibnamefont{Levin}},
  \bibinfo{journal}{Phys. Rev. B} \textbf{\bibinfo{volume}{70}},
  \bibinfo{pages}{064515} (\bibinfo{year}{2004}), \bibinfo{note}{and references
  therein}.

\bibitem[{\citenamefont{Heine et~al.}(1999)\citenamefont{Heine, Lang, Rössler,
  Pedarnig, and Bäuerle}}]{Heine99}
\bibinfo{author}{\bibfnamefont{G.}~\bibnamefont{Heine}},
  \bibinfo{author}{\bibfnamefont{W.}~\bibnamefont{Lang}},
  \bibinfo{author}{\bibfnamefont{R.}~\bibnamefont{Rössler}},
  \bibinfo{author}{\bibfnamefont{J.~D.} \bibnamefont{Pedarnig}},
  \bibnamefont{and} \bibinfo{author}{\bibfnamefont{D.}~\bibnamefont{Bäuerle}},
  \bibinfo{journal}{J. Low Temp. Phys.} \textbf{\bibinfo{volume}{117}},
  \bibinfo{pages}{1265} (\bibinfo{year}{1999}).

\bibitem[{\citenamefont{Dorsey}(1991)}]{Dorsey91}
\bibinfo{author}{\bibfnamefont{A.~T.} \bibnamefont{Dorsey}},
  \bibinfo{journal}{Phys. Rev. B} \textbf{\bibinfo{volume}{43}},
  \bibinfo{pages}{7575} (\bibinfo{year}{1991}).

\bibitem[{\citenamefont{Gauzzi and Pavuna}(1995)}]{Gauzzi95}
\bibinfo{author}{\bibfnamefont{A.}~\bibnamefont{Gauzzi}} \bibnamefont{and}
  \bibinfo{author}{\bibfnamefont{D.}~\bibnamefont{Pavuna}},
  \bibinfo{journal}{Phys. Rev. B} \textbf{\bibinfo{volume}{51}},
  \bibinfo{pages}{15420} (\bibinfo{year}{1995}).

\bibitem[{\citenamefont{Carballeira et~al.}(2001)\citenamefont{Carballeira,
  Curràs, Viña, Veira, Ramallo, and Vidal}}]{Carballeira01}
\bibinfo{author}{\bibfnamefont{C.}~\bibnamefont{Carballeira}},
  \bibinfo{author}{\bibfnamefont{S.~R.} \bibnamefont{Curràs}},
  \bibinfo{author}{\bibfnamefont{J.}~\bibnamefont{Viña}},
  \bibinfo{author}{\bibfnamefont{J.~A.} \bibnamefont{Veira}},
  \bibinfo{author}{\bibfnamefont{M.~V.} \bibnamefont{Ramallo}},
  \bibnamefont{and} \bibinfo{author}{\bibfnamefont{F.}~\bibnamefont{Vidal}},
  \bibinfo{journal}{Phys. Rev. B} \textbf{\bibinfo{volume}{63}},
  \bibinfo{pages}{144515} (\bibinfo{year}{2001}), \bibinfo{note}{and references
  therein}.

\bibitem[{\citenamefont{Blatter et~al.}(1992)\citenamefont{Blatter,
  Geshkenbein, and Larkin}}]{Blatter92}
\bibinfo{author}{\bibfnamefont{G.}~\bibnamefont{Blatter}},
  \bibinfo{author}{\bibfnamefont{V.~B.} \bibnamefont{Geshkenbein}},
  \bibnamefont{and} \bibinfo{author}{\bibfnamefont{A.~I.}
  \bibnamefont{Larkin}}, \bibinfo{journal}{Phys. Rev. Lett.}
  \textbf{\bibinfo{volume}{68}}, \bibinfo{pages}{875} (\bibinfo{year}{1992}).

\end{thebibliography}

\end{document}